\documentclass[12pt,reqno]{article}
\pdfoutput=1
\usepackage{amsmath,hyperref,amssymb}
\usepackage{subfigure,graphicx,url}

\newcommand{\bea}{\begin{eqnarray}}
\newcommand{\eea}{\end{eqnarray}}
\newcommand{\be}{\begin{equation}}
\newcommand{\ee}{\end{equation}}
\newcommand{\mc}{\mathcal}
\newcommand{\Tr}{\text{Tr}}
\newcommand{\tr}{\text{tr}}
\renewcommand{\t}{\tilde}

\allowdisplaybreaks \numberwithin{equation}{section}

\DeclareSymbolFont{AMSa}{U}{msa}{m}{n}
\DeclareSymbolFont{AMSb}{U}{msb}{m}{n}
\DeclareMathSymbol{\fieldR}{\mathalpha}{AMSb}{"52}

\makeatletter
\renewcommand\section{\@startsection {section}{1}{\z@}%
                                 {-3.5ex \@plus -1ex \@minus -.2ex}
                                   {2.3ex \@plus.2ex}%
                                   {\normalfont\large\bfseries}}
\renewcommand\subsection{\@startsection{subsection}{2}{\z@}%
                                   {-3.25ex\@plus -1ex \@minus -.2ex}%
                                     {1.5ex \@plus .2ex}%
                                     {\normalfont\bfseries}}
\renewcommand\subsubsection{\@startsection{subsubsection}{3}{\z@}%
                                   {-3.25ex\@plus -1ex \@minus -.2ex}%
                                     {1.5ex \@plus .2ex}%
                                     {\normalfont\itshape}}
\makeatother

\begin{document}

\begin{flushright} \small
SU-ITP-11/49\\
SLAC-PUB-14648
\end{flushright}
\bigskip
\begin{center}
 {\LARGE \bf
A maximally supersymmetric Kondo model
 }\\[5mm]
 \

{\bf Sarah Harrison, Shamit Kachru, Gonzalo Torroba}   \\[3mm]
 {\small\slshape
 SITP, Department of Physics,
 Stanford University \\
 Stanford, CA 94305-4060, USA \\
\medskip
 Department of Particle Physics and Astrophysics, SLAC\\
 Menlo Park, CA 94309, USA
 \\
\medskip
 {\upshape\ttfamily sarharr@stanford.edu, skachru@stanford.edu, torrobag@slac.stanford.edu}
\\[3mm]}
\end{center}

\

\vspace{5mm}  \centerline{\bfseries Abstract}
\medskip

We study the maximally supersymmetric Kondo model obtained by adding a fermionic impurity to $\mc N=4$ supersymmetric Yang-Mills theory. While the original Kondo problem describes a defect interacting with a free Fermi liquid of itinerant electrons, here the ambient theory is an interacting CFT, and this introduces qualitatively new features into the system. The model arises in string theory by considering the intersection of a stack of $M$ D5-branes with a stack of $N$ D3-branes, at a point in the D3 worldvolume. We analyze the theory holographically, and propose a dictionary between the Kondo problem and antisymmetric Wilson loops in $\mc N=4$ SYM. We perform an explicit calculation of the D5 fluctuations in the D3 geometry and determine the spectrum of defect operators. This establishes the stability of the Kondo fixed point together with its basic thermodynamic properties. Known supergravity solutions for Wilson loops allow us to go beyond the probe approximation: the D5s disappear and
are replaced by three-form flux piercing a new topologically non-trivial $S^3$ in the corrected geometry. This describes the Kondo model in terms of a geometric transition. A dual matrix model reflects the basic properties of the corrected gravity solution in its eigenvalue distribution.

\thispagestyle{empty}

\bigskip
\newpage

\tableofcontents


\section{Introduction}

The Kondo model has been a rich subject of study for almost 50 years.  In a certain sense, it was the first system to exhibit flow from an asymptotically free UV fixed
point to a strongly coupled IR regime.   It has also been a playground for early renormalization group studies, exact conformal field theory techniques, and large N methods
in field theory.
Its generalization to the Kondo lattice system yields models which may be of great relevance in understanding non-Fermi liquid behavior in the heavy
fermion metals.  For reviews with extensive references, see \cite{Ludwig,Affleck}.

A supersymmetric variant of the Kondo system, which may be useful in attempts to apply AdS/CFT to condensed matter systems, was described in
\cite{KKYone} (see also \cite{KKYtwo,Muck,UMich} for related work).  The analogue of the Kondo defect spin arises at an intersection of $M$ D5-branes with a stack of $N$ D3-branes. At large 't Hooft coupling
for the D3-brane field theory, the D5s appear as probes in an $AdS_5 \times S^5$ geometry, wrapping an $AdS_2 \times S^4$ submanifold.
They describe fundamental defects interacting with the ambient $SU(N)$ ${\cal N}=4$ supersymmetric gauge theory (which provides the analogue of
the ``itinerant electrons'' in the Kondo model). Unlike the original Kondo model, in our case the ambient field theory is already at a nontrivial interacting fixed point, giving rise to qualitatively new features that we will explore here.

In this paper we continue the study of this maximally supersymmetric Kondo system, extending the results of \cite{KKYone} in several directions. We will make explicit the correspondence between Kondo impurities and half BPS Wilson loops of $\mc N=4$ SYM, and this will allow us to extract new results at strong coupling. Some of the main questions about the Kondo model involve 
the physics reflected in the defect entropy, specific heat, and magnetic susceptibility.  In the latter two cases, the first non-trivial contribution involves knowledge
of the leading irrelevant operator (invariant under any relevant global symmetries, like spin and channel symmetries) at the IR fixed point
\cite{Ludwig,Affleck}.   So to determine the basic physics, one must find the spectrum of operators in the defect theory, which maps in the gravity
regime to a classification of KK modes on the probe D5s.  The complete spectrum and map to the dual field theory has not yet been determined;
we complete this task here.  Our results reveal that the superconformal algebra which governs the fixed point places powerful constraints on the
defect thermodynamics (and can even strongly limit the effects of inter-defect interactions, in models with multiple D5 branes).

While gauge/gravity duality immediately sums the effects of the large $N$ gauge theory on the defects (through the details of the geometry wrapped
by the D5-brane probes), the leading approximation does not include the backreaction of the defect flavors on the ${\cal N}=4$ supersymmetric field theory.  
To determine this at strong 't Hooft coupling, one must find a corrected solution to the supergravity equations, including the non-trivial effects of the D5-brane stress-energy
sources instead of simply treating the D5s as probes.  These solutions are already known, from earlier work of 
D'Hoker, Estes and Gutperle \cite{DEG}. We describe them explicitly here and apply their results to the analysis of our supersymmetric Kondo model, now including backreaction. The D5-branes induce a geometric transition, changing the global topology of the solution, and
replacing themselves with three-form flux piercing a new non-trivial $S^3$ in the corrected geometry. 
We also discuss how the physics of these backreacted solutions is captured by a matrix model \cite{Yamaguchi}, which may
be useful in carrying out further field theory computations.

In addition to their intrinsic interest, the questions we investigate here are well motivated because understanding the physics of the single-defect
model is the logical starting point for a more ambitious program of studying gravity duals to lattices of such defects, analogous to Kondo lattice systems.
Kondo  lattice models of defect spins interacting with a bulk Fermi liquid are thought to give rise to the fascinating non-Fermi liquid phases in the heavy
fermion systems (which in turn may be closely related to the ``strange metal" phase of the cuprate high temperature superconductors).  
The papers \cite{AdSRN} initiated a holographic approach to non-Fermi liquids using special properties of the $AdS_2$ geometry that emerges in
the near-horizon limit of the Reissner-Nordstrom black brane.  Lattice models of defect flavors provide an alternative method of obtaining the geometries
that can give rise to non-Fermi liquid behavior \cite{KKYtwo,Jensenetal}, with some important differences from the Reissner-Nordstrom case; the $AdS_2$ geometries  now arise on the probe world-volumes themselves.  General discussions of why Kondo lattice models could be relevant to the problem of strange metals from a field-theoretic perspective appear in \cite{Sachdevone,Sachdevtwo}.

The organization of this paper is as follows.  
In \S2, we describe basic field theoretic aspects of the Kondo model.  
In \S3, we review and extend the work of \cite{KKYone}, discussing the simplest results which are visible from a probe approximation.  In \S4, 
we turn to a determination of the full spectrum of operators in the defect theory; the new defect-localized operators are found by performing a KK reduction
on the wrapped probe branes.  The extended superconformal algebra present in this theory plays a powerful role in organizing the results.
With this list of operators in hand, in \S5 we turn to its implications for the basic questions in Kondo physics -- questions of the IR stability of the Kondo fixed
point, and the leading non-trivial contributions to defect specific heat and magnetic susceptibility.
Finally, in \S6, we describe the supergravity solutions which incorporate backreaction of the defects on the bulk, as well as a dual matrix model
which reproduces the main features of the supergravity solutions in its eigenvalue distribution.
\S7 contains our concluding remarks about promising future directions.  Some unpalatable details of the computation of the
spectrum in \S4 have been relegated to the appendix.

\section{Field theory analysis}\label{sec:QFT}

This first section is devoted to a field theory analysis of the Kondo problem and its maximally supersymmetric version. After reviewing the original multichannel Kondo model, we describe generalizations based on interacting conformal field theories (CFTs), with a focus on the supersymmetric case that arises from $\mc N=4 $ SYM coupled to a defect fermion. The quantum theory is studied in two complimentary ways. First we integrate out the interactions with the ambient theory and obtain an effective quantum mechanics description of the impurity at low energies. This provides a simple way of characterizing the RG evolution of the Kondo coupling and of evaluating correlation functions of the defect fermions. On the other hand, we may also integrate out the defect fermions and derive a new ambient theory where the impurity has been replaced by certain boundary conditions on the bulk fields. Their equivalence will reveal a very useful correspondence between Kondo impurities and Wilson loops.

\subsection{Multichannel Kondo model}\label{subsec:multichannel}

The multichannel Kondo model is described by the Hamiltonian
\be\label{eq:H1}
H = \sum_{\vec p,\,i,\,\alpha}\,\epsilon(\vec p) \psi^\dag_{\vec p\,i \alpha}\psi_{\vec p\,i \alpha}+ J   \sum_{\vec p\,\vec{p'}\,i \alpha \beta}S_{\alpha \beta}\, \psi^\dag_{\vec p \,i \alpha} \psi_{\vec{p'}\,i\beta}
\ee
where $\psi^\dag_{\vec p\,i \alpha}$ is the creation operator for a conduction electron with channel index $i=1,\ldots,K$, $SU(N)$ spin $\alpha$, and energy
\be
\epsilon(\vec p)= \frac{p^2}{2m}-\epsilon_F\,.
\ee
The interaction is antiferromagnetic, $J>0$. These ``ambient'' electrons interact with a spin $S_{\alpha \beta}$ localized at the origin and transforming under a representation R of $SU(N)$. The Fourier transform of this delta-function interaction gives rise to the sum over momenta in the second term of (\ref{eq:H1}). The $SU(N)$ generalizes spin, and this extension will be important for us because the system that we will study has $N=K$ (large).

Our analysis will focus on the antisymmetric representation with $k$ indices, in which case the impurity spin $S$ can be described in terms of auxiliary fermions $\chi_\alpha$ with total fermion number $k$:
\be
S_{\alpha \beta} = \chi^\dag_\alpha \chi_\beta - \frac{k}{N} \delta_{\alpha \beta}\;\;,\;\;\sum_{\alpha=1}^N \chi^\dag_\alpha \chi_\alpha = k\,.
\ee

The original version of the model corresponds to one channel (i.e. $K=1$) with $SU(2)$ spin ($N=2$), interacting with a single impurity fermion. Integrating out the conduction electrons up to some cutoff of order of the Fermi momentum shows that at low temperatures the Kondo coupling flows to $J \to \infty$. This means that the Hamiltonian (\ref{eq:H1}) is minimized by binding one electron with the impurity to form a spin singlet (recall that $J>0$). Therefore, the impurity spin is completely screened and disappears from the low energy theory. We are left with a theory of free electrons, but we have to impose the boundary condition $\psi=0$ at the origin. This phenomenon will have an analog in the gravity dual, in terms of a geometric transition where the D5 branes that give rise to the impurity disappear and are replaced by their flux.

In the multichannel $K \ge 2$ case, the ground state of the strong coupling point $J \to \infty$ is a Young tableau with $K-1$ columns and $N-k$ lines; this has larger dimension than the degeneracy at zero coupling, given by $N!/k!(N-k)!$ (see e.g. \cite{kotliar}). Therefore, a RG flow from weak to strong coupling is not possible, and instead an intermediate fixed point is expected. This is known as ``overscreening'', and is not described by Fermi-liquid theory. Our supersymmetric model will correspond to this regime, and we will study the intermediate fixed point using AdS/CFT.

Let us now summarize some of the results on the Kondo model, obtained using CFT and large $N$ techniques. In the approach of Affleck and Ludwig (see \cite{Affleck,Ludwig} for review and references), the system is first mapped to a 2d theory by performing an s-wave reduction. The ambient CFT is decomposed into spin, flavor and charge degrees of freedom, and is described by a Kac-Moody algebra $\widehat{SU}_K(N)_s \otimes \widehat{SU}_N(K)_f \otimes \widehat{U}(1)_c$. At the intermediate fixed point, the local impurity $S$ disappears, becoming part of the spin current. Thus, its correlator is fixed by conformal invariance,
\be\label{eq:Deltas}
\langle S(0) S(t) \rangle \sim \frac{1}{t^{2\Delta_s}}\;\;,\;\;\Delta_s = \frac{N}{N+K}\,.
\ee
The whole effect of the impurity is encoded into constraints that determine the number of times a given state occurs in the spectrum. According to the ``fusion rule'', the spectrum is determined by acting on a given primary operator with the spin current in the representation $R$ of the impurity spin. The flavor and charge sectors are not modified. These constraints give rise to a conformal field theory with boundary conditions that respect conformal invariance --a boundary CFT.

The approach of integrating out the ambient electrons to obtain an effective theory on the defect has also been followed in the literature; see e.g. \cite{kotliar}. The saddle point equations for the defect fields can be solved at large $N$, leading to results consistent with the boundary CFT predictions.

We will then compare these to the predictions from the gravity side. Particularly important is the impurity entropy (or g-function), which is defined as
\be\label{eq:Simp}
\log g= \mc S_\text{imp} \equiv \text{lim}_{T \to 0} \,\text{lim}_{V \to \infty} \left[\mc S(T) - \mc S_\text{ambient}(T) \right]
\ee
where $\mc S_\text{ambient}$ is the contribution proportional to the volume $V$. The g-function is the analogue of the c-function for theories with boundary. In the multichannel model,
\be\label{eq:Simp2}
\mc S_\text{imp}= \log\,\prod_{n=1}^k \frac{\sin\,\frac{\pi(N+1-n)}{N+K}}{\sin\,\frac{\pi n}{N+K}}\,.
\ee

\subsection{Supersymmetric Kondo model}\label{subsec:general}

The Kondo fixed point is naturally analyzed in terms of a $1+1$-dimensional CFT, obtained by performing an s-wave reduction
\be
\psi(\vec p) = \frac{1}{\sqrt{4\pi} p} \psi_0(p) + \textrm{higher harmonics}
\ee
and keeping only the spherically symmetric wavefunctions. Linearizing around the Fermi surface and using the doubling trick obtains
\be\label{eq:H2}
H= \int_{- \infty}^{+\infty} dx\, \psi_L^\dag \left(i\partial_x+  J \delta(x)\,S\right)\psi_L\,.
\ee
The model reduces to a 2d CFT coupled to degrees of freedom living on the boundary. A natural extension is to couple an interacting CFT to a localized impurity,
\be
S = S_{CFT}+ \int dt\,\bar \chi \left(i \partial_t + \mathcal O(x=0) \right) \chi\,,
\ee
where $\mc O$ is some operator of the CFT, and the impurity is at the origin $x=0$. We refer the reader to the review \cite{Sachdevtwo} and references \cite{sachdevthree} for works on this approach in the condensed matter literature; in particular the latter papers study Kondo defects coupled to CFTs, albeit simpler CFTs than the ${\cal N}=4$ super Yang-Mills theory.

Our AdS/CFT techniques will be closest to this approach, albeit with additional supersymmetry and in the large N limit. We couple $\mc N=4$ supersymmetric $SU(N)$ gauge theory (the ambient fields) to $M$ $(0+1)$-dimensional fermions $\chi$, with total fermionic number $k$ (the ``defect''). The action reads
\be\label{eq:S1}
S= S_{\mc N=4}+ \int dt\,i \left[\bar \chi_i^I \partial_t \chi^i_I+ \bar\chi_i^I (T^A)^i_j \left(A_0(t,0)_A+n_a \phi^a(t,0)_A \right) \chi^j_I\right]\,,
\ee
where $i,j= 1, \ldots, N$, $I= 1, \ldots, M$. Here $T^A$ are the generators in the adjoint of $SU(N)$, and $n^a$ is a unit vector in $\mathbb R^6$. We also need to impose the constraint $\sum_i \bar \chi_i \chi_i=k$, which fixes the fermion number. It will be convenient to impose this using a Lagrange multiplier $M \times M$ matrix $\tilde A_0$, and adding
\be\label{eq:multiplier}
S_k \equiv \int dt\left(\bar \chi^I_i (\tilde A_0)_I^J \chi_J^i - k (\tilde A_0)_I^I \right)\,,
\ee
to the action (\ref{eq:S1}). In the brane construction below, $\tilde A_0$ arises from the D5 gauge field.

In the conventions of (\ref{eq:S1}), the kinetic term for the gauge field and scalars is proportional to $1/g_{YM}^2$, so the coupling between the impurity and bulk degrees of freedom is proportional to the gauge coupling $g_{YM}$. The $\mc N=4$ theory preserves 16 supercharges $\epsilon$; half of these are broken by the defect,
\be
\gamma_0 n^a\gamma_a \epsilon = \epsilon\,,
\ee
where $\gamma_M$ are the 10d Dirac matrices. The defect fermions are singlets under the preserved supersymmetries. 
The large amount of supersymmetry, which fixes the Kondo coupling in terms of $g_{YM}$, has important consequences on the quantum theory. We will argue shortly that the defect action does not receive quantum corrections; furthermore, having a fixed Kondo coupling restricts the spectrum of defect operators, as will be seen in \S \ref{sec:excitations}.

Comparing with the multichannel model (\ref{eq:H1}), we see that our theory corresponds to $K=N$. This particular case turns out to be quite interesting on its own, because matrix model techniques can be applied to the Kondo problem, as we discuss in more detail in \S \ref{sec:backreacted}. On the other hand, the defect fermions $\chi_i^I$ have an additional $U(M)$ flavor number that is absent from (\ref{eq:H1}). This allows for more general (antisymmetric) representations for the spin impurity.

In \S \ref{sec:branes} we will review how this field theory arises from $N$ D3 branes intersecting $M$ D5 branes (with worldvolume flux $k$) at a point. Taking the limit of large $N$ and  strong coupling will then allow us to solve for many properties of the theory using the gravity dual. Before moving in this direction, in the rest of the section we discuss basic properties of the quantum theory at weak coupling.

\subsection{Quantum effects from the ambient CFT}\label{subsec:quantum}

Let us integrate out the ambient fields to obtain an effective $0+1$ dimensional theory. We start from (\ref{eq:S1}) at weak coupling and zero temperature, and integrate out $A_0 + n_a \phi^a$. The effective action on the defect will be of the form
\be
S_{eff}= \int dt \,i\,\bar\chi_i^{I} \partial_t \chi^i_I + \frac{g_{YM}^2}{8\pi^2}\int dt\, dt'\,D(t-t')\, \bar\chi^{I}_i(t) \chi_I^j(t) \bar\chi^{J}(t')_j \chi_J^i(t')+S_k \,,
\ee
where the kernel $D(t-t')$ is generated by the scalar and gluon propagators,
\be\label{eq:D1}
\langle A_0(t)_A A_0(t')_B+ n_a n_b \phi^a(t)_A \phi^b(t')_B\rangle = \frac{g_{YM}^2}{4\pi^2}\,\delta_{AB} D(t-t')\,.
\ee
$A, B$ are indices in the adjoint, while $a,b$ run over the six scalars of $\mc N=4$ SYM. Also $S_k$ was defined in (\ref{eq:multiplier}) and determines the fermion number.

To continue, we need the explicit form of these propagators, which at tree level (and in Feynman gauge) is
\bea\label{eq:prop}
\langle \phi_A^a(x) \phi_B^b(x') \rangle &=& \frac{g_{YM}^2}{4\pi^2} \frac{\delta_{AB} \delta^{ab}}{(x-x')^2}\nonumber\\
\langle A_\mu^A(x) A_\mu^B(x') \rangle &=& \frac{g_{YM}^2}{4\pi^2} \frac{\eta_{\mu\nu} \delta^{AB}}{(x-x')^2}\,,
\eea
where $\eta_{\mu\nu}= \text{diag} (-1,1,1,1)$. Replacing (\ref{eq:prop}) into (\ref{eq:D1}) reveals something quite striking: the contributions from $A_0$ and $\phi$ exactly cancel, giving $D(t-t')=0$! In fact, this is also true to all orders in the gauge coupling \cite{Semenoff,Gross,Pestun}. We conclude that the defect theory is not renormalized at zero temperature. This rather special behavior is due to the large amount of supersymmetry, which protects the defect action against quantum corrections and fixes the Kondo interaction in terms of $g_{YM}$.

Next we put the theory at finite temperature  $T=1/\beta$ and analytically continue to imaginary time $\tau=it$. This breaks supersymmetry and we expect nontrivial quantum corrections.\footnote{A recent related discussion has appeared in \cite{Muck}, where the behavior of $D(t)$ was conjectured based on comparisons with the D5 probe limit. Here we derive $D(t)$ directly from the field theory, thus verifying the assumption of \cite{Muck} and obtaining nontrivial agreement with the gravity prediction.} The propagators at finite temperature can be obtained from the zero temperature limit (\ref{eq:prop}) by an inversion that maps the line (the time direction) to a circle. The important point here is that while the scalar propagator is invariant, the gluon contribution is not. The gluon propagator is just changed by a total derivative \cite{Gross} or, equivalently, a constant contribution to $D(\tau-\tau')$ after replacing into the effective action. On dimensional grounds, $D \sim 1/\beta^2$.\footnote{This may also be calculated from the circular Wilson loop, as explained e.g. in \cite{Gross}.} In contrast, the ambient fermions in the Kondo model of \S \ref{subsec:multichannel} have a propagator that scales like $1/\tau$.

In summary, the defect theory at finite temperature becomes
\be\label{eq:Seff1}
S_{eff}= \int d\tau \,\bar\chi_i^{I} \partial_\tau \chi^i_I + \frac{g_{YM}^2}{32 \beta^2}\int d\tau\, d\tau'\, \bar\chi^{I}_i(\tau) \chi_I^j(\tau) \bar\chi^{J}(\tau')_j \chi_J^i(\tau')+S_k\,.
\ee
Equipped with this action, we are now ready to calculate the fermion correlator and impurity entropy, following \cite{kotliar, Muck}. From (\ref{eq:multiplier}), $\t A_0$ plays the role of the chemical potential for the defect fermions; let's denote its expectation value by
\be
\bar \mu = \left\langle \frac{1}{\beta} \int_0^\beta d\tau \,i\t A_0 \right\rangle\,.
\ee
The quartic interaction gives a contribution to the self-energy $\Sigma \propto D(\tau) G(\tau)$, so at large $N$ the Schwinger-Dyson equation for the two-point function becomes
\be
G(\omega) = \frac{1}{\omega-\bar \mu - \frac{\lambda}{16 \beta^2} G(\omega)}
\ee
where the 't Hooft coupling $\lambda = N g_{YM}^2$ comes from the quartic interaction and the $N$-component fermion in the loop. The solution is
\be\label{eq:G}
G(\omega)= \frac{1}{2} \left(\frac{\lambda}{16 \beta^2} \right)^{-1} \left(\omega - \bar \mu - \left[(\omega - \bar \mu)^2 - 4  \frac{\lambda}{16 \beta^2}\right]^{1/2} \right)\,.
\ee

The result (\ref{eq:G}) determines the basic properties of the system. The chemical potential is fixed in terms of the fermion number $k$ by $G(\tau \to 0) = k/N$. At large 't Hooft coupling this yields \cite{Muck}
\be\label{eq:barmu}
\frac{k}{N}= \frac{1}{2}- \frac{1}{\pi} \arcsin\,\hat \mu - \frac{1}{\pi} \hat \mu \sqrt{1-\hat \mu^2}\;,\; \hat \mu \equiv \frac{2 \beta}{\sqrt{\lambda}} \bar \mu\,.
\ee
In particular, this implies that the chemical potential increases with $\lambda$ as $\bar \mu \propto \sqrt{\lambda}/\beta$. Now, recall that $\bar \mu$ is related to the defect free energy by $\bar \mu = - \partial F_\text{defect}/\partial k$; deriving once more with respect to the temperature obtains
\be
\frac{\partial \mc S_\text{imp}}{\partial k}= - \frac{\partial \bar \mu}{\partial T}\,.
\ee
Using the result (\ref{eq:barmu}) we thus arrive to the impurity entropy
\be\label{eq:qftSimp}
\mc S_\text{imp}= \frac{N \sqrt{\lambda}}{3\pi} (1- \hat \mu^2)^{3/2}\,.
\ee
The same result will be obtained from the gravity side, with $\hat \mu = \cos \theta_k$ being related to the angle that defines the embedding of the D5 branes inside the $S^5$. Notice that the impurity entropy increases with the 't Hooft coupling --an effect that is absent in the original Kondo model. From the field theory calculation, this is due to the fact that the chemical potential increases like $\sqrt \lambda$.

\subsection{Impurities as Wilson loops}\label{subsec:Kondo-Wilson}

We can also derive an effective theory for the bulk fields by integrating out the defect fermions, which can be done exactly. This produces a Wilson loop insertion,
\be
W_R= \Tr_R P \exp\left(i \int dt (A_0+ n^a \phi_a) \right)
\ee
in a representation $R$ of the gauge group. This establishes a correspondence between the supersymmetric Kondo model and Wilson loops in $\mc N=4$ SYM. In the rest of the paper we will develop this dictionary in detail, and use exact results on Wilson loops to understand the strong coupling behavior of the Kondo model.\footnote{A similar relation between Kondo-CFT problems and Wilson loops was discussed in the condensed matter literature in e.g. \cite{SachdevWilson}.}

Let us review how the Wilson loop is generated in the simplest case $M=1$; we follow \cite{Gomis}. It is convenient to choose a gauge where $A_0+ n^a \phi_a$ has constant eigenvalues $m_1, \ldots , m_N$.
The equation of motion for the defect fermions is
\be
(i \partial_t + m_i) \chi_i=0\;\;,\;\;i=1, \ldots, N\,.
\ee
We have a Fock space of $N$ fermions, each with energy $m_i$, and in the partition function we only need to sum over states with $k$ fermions. The partition function of the defect fermions then becomes
\be\label{eq:Zdefect}
Z_\text{defect}= \sum_{i_1<i_2< \ldots< i_k} e^{i \int dt \,m_{i_1}} \ldots e^{i \int dt \,m_{i_k}}\,,
\ee
where an infrared regulator is assumed. Reintroducing the combination $A_0+n^a \phi_a$, (\ref{eq:Zdefect}) is recognized as the trace of the Wilson line in the $k$-th antisymmetric representation of $SU(N)$,
\be
\sum_{i_1<i_2< \ldots< i_k} e^{i \int dt \,m_{i_1}} \ldots e^{i \int dt \,m_{i_k}}= \Tr_{A_k}\,P\,\exp\left(i \int dt (A_0+n^a \phi_a) \right)\,.
\ee
This shows that integrating out the degrees of freedom that live on the defect produces a supersymmetric Wilson loop insertion in the $\mc N=4$ theory.

In the case $M>1$ we have $N M$ fermions with total fermion number $k$. This allows for general representations for the Wilson line, where the Young tableaux has at most $N$ rows and $M$ columns, and the number of boxes is determined by $k$. (For $M=1$ this simply gives $k$ rows). Now the matrix structure of $\t A_0$ needs to be specified as well; choosing
\be
\t A_0 =\text{diag}(\Omega_1, \ldots, \Omega_M)\,,
\ee
obtains an equation of motion
\be
(i \partial_t+ m_i+\Omega_I) \chi_i^I=0\,.
\ee
So there are $NM$ fermions, with energy $m_i+\Omega_I$ and total fermion number $k$. The chemical potential $\Omega_I$ determines the number of boxes in the $I$-th column as the value of the propagator $\langle \bar \chi_I(\tau) \chi^J(0)\rangle$ as $\tau \to 0$.

\section{D3-D5 system in the probe approximation}\label{sec:branes}

Having understood the basic properties of the field theory with a quantum impurity at weak coupling, in what follows we will study the Kondo model at large 't Hooft coupling using AdS/CFT. We begin by reviewing the realization of quantum impurities using intersecting D3 and D5 branes. Next we study the model at large 't Hooft coupling and large $N$, but keeping the number of D5 branes $M$ fixed. As explained in \cite{KKYone}, this is equivalent to having D5 branes embedded in $AdS_5 \times S^5$ with worldvolume flux. From the field theory perspective, this is a quenched approximation where loops of defect fields are neglected. After reviewing the gravity description, we calculate the g-function and compare with the nonsupersymmetric Kondo model of \S \ref{subsec:multichannel}.

\subsection{D-brane description of the Kondo model}

The brane configuration is given by\footnote{We refer the reader to \cite{Yamaguchi, Gomis} for detailed descriptions of this system. }
\begin{center}
\begin{tabular}{c|cccc|cccccc}
    & $0$ & $1$&$2$&$3$&$4$&$5$&$6$&$7$&$8$&$9$ \\
\hline
&&&&&&&&&\\[-12pt]
$N\;\;D3$ &$\times$ & $\times$&$\times$ &$\times$ & & & & & & \\
$M\;\;D5$ &$\times$ & & & & $\times$ &$\times$ &$\times$ & $\times$&$\times$ & \\
$k\;\;F1$ &$\times$ & & & &  &  &  & &  & $\times$ 
\end{tabular}
\end{center}
This configuration preserves 8 supercharges. The $k$ fundamental strings between the D3 and D5 branes lead to $k$ fermions localized on the D3/D5 intersection.

The full open string action is given by
\be\label{eq:fullS}
S = S_{D3} + S_{D5}+ S_\text{defect}
\ee
where $S_{D3}$ is the action for the $N$ D3-branes (namely $\mc N=4$ SYM), $S_{D5}$ describes the 6d worldvolume theory on the $M$ D5-branes, and $S_\text{defect}$ is the action for the 3-5 strings at the intersection, 
\be\label{eq:Sdefect1}
S_\text{defect}  = \int dt\, \left[i\bar \chi_i^I \partial_t \chi^i_I+ \bar\chi_i^I  \left(A_0(t,0)^i_j+n_a \phi^a(t,0)^i_j \right) \chi^j_I + \bar \chi^I_i (\tilde A_0)_I^J \chi_J^i - k (\tilde A_0)_I^I \right]
\ee
where $i,j= 1, \ldots, N$, $I= 1, \ldots, M$. The fermion $\chi$ is a bifundamental of $SU(N) \times SU(M)$, so it couples to the D3 and D5 gauge fields $A_0$ and $\t A_0$. Supersymmetry requires that it couple also to one linear combination $n_a \phi^a$ of the scalars of the $\mc N=4$ theory; $n^a$ is a unit vector in $\mathbb R^6$. $\t A_0$ also acts as a Lagrange multiplier fixing the fermion number $\bar \chi^I_i \chi_J^i = k \delta^I_J$.

In the decoupling limit for the D3 gauge theory, the 5-5 strings become nondynamical. Then we are left with 4d $\mc N=4$ SYM with gauge group $SU(N)$, coupled to a $0+1$-dimensional fermionic system (\ref{eq:Sdefect1}) with flavor symmetry $U(M)$. This reproduces the supersymmetric Kondo model of \S2.2.

Let us now discuss the symmetries preserved by the defect. A line $(x^0,0,0,0)$ preserves rotations along the line, time translations, dilatations and special conformal transformations along the time direction, $\delta x^0= b^0 x^2$. These generators $(J_{ij}, P_0, D, K_0)$ give $SO(3) \times SL(2, \mathbb R)$; this is the subgroup of the original $SO(4,2)$ conformal group that is left unbroken by a line. On the other hand, the $SO(6)$ R-symmetry that acts on the $\mc N=4$ SYM scalar fields is broken to $SO(5)$ by the coupling $\bar \chi n_a \phi^a \chi$. Thus, the bosonic symmetries preserved by the Wilson loop are $SL(2, \mathbb R) \times SO(3) \times SO(5)$. The defect also breaks half of the supersymmetries of the ambient gauge theory. These combine with the bosonic symmetries to give the supergroup $OSp(4^*|4)$ \cite{Gomis}. This group plays an important role in understanding the dynamics at strong coupling. The operators of lowest dimension correspond generically to short multiplets of $OSp(4^*|4)$, whose dimensions are protected and determined by group theory. We calculate the spectrum of excitations explicitly in \S \ref{sec:excitations}, and use these results in \S \ref{sec:thermo} to study the stability and thermodynamics of the defect theory.

The dynamics of the system is controlled by the 't Hooft coupling $\lambda = g_{YM}^2 N$ and $M/N$. At weak coupling $\lambda \ll 1$ we can perform perturbative field theory calculations, and the results of \S \ref{sec:QFT} apply. On the other hand, at large 't Hooft coupling there is dual gravity description in terms of type IIB supergravity with 3- and 5-form fluxes. When $M/N \ll 1$ the D5 branes can be treated as probes in the $AdS_5 \times S^5$ background generated by the D3 branes. For $1 \ll g_{YM}^2 M \ll g_{YM}^2 N$ the backreaction of the 5-branes should be taken into account as well. We now focus on the probe limit, postponing the analysis of the fully backreacted solution to \S \ref{sec:backreacted}, where we will make contact with the results of \cite{DEG}.

\subsection{Gravity side}\label{subsec:probegravity}

Let us review the gravity description of Wilson loops in the probe limit, where $M/N$ corrections are ignored, and apply it to the Kondo model. As found in \cite{KKYone}, in this approximation the quantum impurity is described holographically by D5 branes embedded in $AdS_5 \times S^5$, with worldvolume flux. In field-theoretic terms, this gives a solution to the Kondo model in the quenched approximation, where loop effects of the defect fermions are ignored. Furthermore, the ambient conformal field theory is very strongly coupled, and its quantum effects are resummed into a gravitational dual. This approximation will allow us to find important quantities such as the defect entropy and thermodynamics controlled by the leading irrelevant operators.

In order to realize the defect symmetries transparently, it is useful to slice $AdS_5 \times S^5$ by
\be\label{eq:AdSfiber1}
ds^2 = R^2 \left(du^2 + \cosh^2 u \,ds_{AdS_2}^2+ \sinh^2 u\,d\Omega_2^2 + d\theta^2 + \sin^2 \theta\,d\Omega_4^2\right)\;\;,
\ee
with $u \ge 0$ and $0\le \theta \le \pi$. The potential $C_4$ that gives rise to the five form flux can be written as \cite{Yamaguchi}
\be\label{eq:C4}
C_4 = R^4 \left[\left(-\frac{u}{2}+ \frac{1}{8} \sinh 4u \right) e^0 e^1 e^4 e^5+ \left(\frac{3}{2} \theta- \sin 2 \theta + \frac{1}{8} \sin 4 \theta \right)  e^6 e^7 e^8 e^9\right]\,,
\ee
where $(e^0, e^1)$ are the vielbeins for $ds_{AdS_2}^2$, $(e^4, e^5)$ are the vielbeins for $d\Omega_2^2$, and $(e^6, e^7, e^8, e^9)$ are those for $d\Omega_4^2$.

The D5-brane worldvolume is $AdS_2 \times S^4$, given by the embedding conditions
\be\label{eq:embeddingD5}
u=0\;\;,\;\;\theta=\theta_k\,.
\ee
As we review shortly, this angle is related to the number of defect fermions $k$ by
\be\label{eq:thetak}
k = \frac{N}{\pi} \left(\theta_k - \frac{1}{2} \sin 2\theta_k \right)\,. 
\ee
This configuration was studied originally by \cite{Paredes}. The bosonic symmetries $SL(2, \mathbb R) \times SO(3) \times SO(5)$ are explicitly realized in the way of writing the metric (\ref{eq:AdSfiber1}). 

Ignoring nonabelian interactions, the D5 branes are described by the DBI action
\be\label{eq:DBI}
S_5 = - T_5 \int d^6 \xi \,\sqrt{-\det(G+F)}+ T_5 \int F \wedge C_4\,,
\ee
with worldvolume metric 
\be\label{eq:g5}
ds_{D5}^2 = R^2(ds_{AdS_2}^2 + \sin^2 \theta_k\,d\Omega_4^2)\,.
\ee
The background value (\ref{eq:C4}) induces a tadpole for $F$ due to the Chern-Simons interaction. As a result, there is nonzero worldvolume flux given by 
\be\label{eq:wvflux}
F = \cos \theta_k\,e^0 \wedge e^1\,,
\ee
and the 
fundamental string charge $k$ is given by $\delta S_{D5}/\delta F_{01}$.

\subsection{Defect free energy and entropy}\label{subsec:gfc}

As a first step towards solving the supersymmetric Kondo model, we now evaluate the boundary free energy and entropy (g-function) in the probe approximation, and compare with the multichannel Kondo model of \S \ref{subsec:multichannel}.

The defect free energy, as a function of temperature $T$, can be computed using the DBI action immersed in an AdS black hole. This calculation is easier to perform in the original Poincare slicing of AdS, where the black hole metric obtains the familiar form
\begin{equation}
ds^2 = -f(r) dt^2 + {dr^2 \over f(r)} + {r^2\over R^2} \left( \sum_{i=1}^3 dx_i^2 \right) + R^2 (d\theta^2 + {\rm sin^2} \theta\, d\Omega_4^2)~,
\end{equation}
where
\begin{equation}
f(r) = {r^2 \over R^2} \left( 1 - {r_+^4 \over r^4} \right)\,.
\end{equation}
The temperature of the field theory is related to the location of the horizon by $T= r_+/ \pi R^2$. The 5-form flux is
\begin{equation}
\label{efffive}
F_5 = -(1 + *)  {4 r^3 \over R^4} (dt \wedge dr \wedge dx_1 \wedge dx_2 \wedge dx_3)\,.
\end{equation}
In these coordinates, the D5 branes are extended along $r$, $t$ and the $S^4$.

After regularizing by subtracting the Euclidean action of the analogous D5 embedded into the pure $AdS_5 \times S^5$ spacetime, the D5 free energy at leading order in $1/N$
is given by \cite{Hartnoll:2006hr}
\begin{equation}\label{eq:FD5}
F_\text{defect} = -\sqrt{\lambda}\,\frac{\sin^3 \theta_k} { 3\pi} N  T ~,
\end{equation}
where $\lambda$ is the 't Hooft coupling $\lambda = g_{YM}^2 N$. Ignoring nonabelian interactions, for $M$ D5 branes 
the total free energy is simply $M$ times the above result. Recalling the field theory definition (\ref{eq:Simp}) of impurity entropy and that $\mc S = - \partial F / \partial T$, from (\ref{eq:FD5}) we obtain
\be\label{eq:Simpprobe}
\log g = \mc S_\text{imp} = \sqrt{\lambda}\,\frac{\sin^3 \theta_k} { 3\pi} M N\,,
\ee
in agreement with the field theory result (\ref{eq:qftSimp}). We note that this coincides with the expectation value of the circular Wilson loop, up to a factor of 2 from mapping the Polyakov to the circular loop \cite{Muck}.

This expression, which is valid at large 't Hooft coupling and when $N/M \ll 1$, is independent of temperature. The result  applies to any nonzero (even very small) temperatures and is dictated by conformal invariance. On the other hand,  CFT techniques also determine the defect entropy of the overscreened Kondo model, given in (\ref{eq:Simp2}). At large $N$, this can be approximated by \cite{kotliar}
\be\label{eq:Simpmulti}
\mc S_\text{imp} = \frac{2}{\pi} MN\left[f\left(\frac{\pi}{1+K/N} \right)-f\left(\frac{\pi}{1+K/N}(1-k/N) \right)-f\left(\frac{\pi}{1+K/N}k/N\right) \right]
\ee
where $f(x) = \int_0^x du \,\log \sin u$ and recall that $K$ is the number of channels of the ambient electrons.

A crucial difference between both expressions is that (\ref{eq:Simpprobe}) increases with the 't Hooft coupling, which fixes the strength of interactions in the ambient CFT. However, the multichannel Kondo model of \S \ref{subsec:multichannel} is based on a gas of noninteracting electrons, and this effect is absent. Physically, the increase of $\log g$ with $\sqrt \lambda$ in our supersymmetric model is due to the fact that the chemical potential for the defect fermions is proportional to $\sqrt \lambda$. This may be understood directly in field theory terms, as we found in \S \ref{subsec:quantum}. It implies that the $g$-function increases under a marginal deformation of the ambient theory by the operator $F_{\mu\nu}^2$ that changes $g_{YM}$ --a property that could also be understood using the approach of \cite{Green:2007wr}.

\begin{figure}[htb]
\begin{center}     
\includegraphics[width=.52\textwidth]{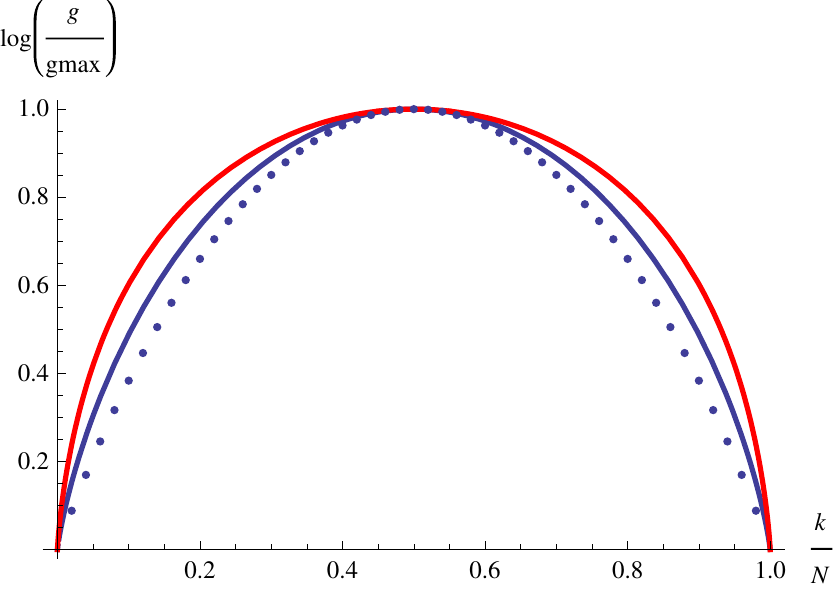}
\caption{\small{Impurity entropy as a function of $k/N$ for the supersymmetric model (dotted curve) and nonsupersymmetric multichannel model with number of channels $K/N=1$ (blue) and $K/N=0.1$ (red).}}
\label{fig:gfc}
\end{center}
\end{figure}

To continue comparing the two answers, note that both g-functions become largest when $k/N=1/2$, corresponding to having particle-hole symmetry $\chi^\dag \leftrightarrow \chi$. Normalizing the g-functions so that their value is one at the maximum, Figure \ref{fig:gfc} shows $\log g$ as a function of $k/N$ by numerically solving (\ref{eq:Simpprobe}) (dotted line) and from (\ref{eq:Simpmulti}) for the values $K/N=1$ (blue curve) and $K/N=1/10$ (red curve). In particular, this reveals that the two entropies are most similar when $K=N$, namely when the ambient electrons are $N \times N$ matrices --as we would expect when comparing to the $\mc N=4$ model.

It is also useful to compare the two behaviors in the regime $k/N \ll 1$. Expanding (\ref{eq:Simpprobe}) in powers of $k/N$ obtains
\be
\log g = \frac{1}{2} kM \sqrt{\lambda} \left[1- \frac{3}{10}\left(\frac{3\pi k}{2N} \right)^{2/3}- \frac{3}{280} \left(\frac{3\pi k}{2N} \right)^{4/3} + \ldots \right]\,.
\ee
This intriguing nonanalytic dependence was already noticed by \cite{Hartnoll:2006hr,Yamaguchi}. Interestingly, the nonsupersymmetric CFT result also shows a nonanalytic dependence, albeit a logarithmic one:
\be
\log g = k M\left[1- \log\left(\frac{\pi}{2} \frac{k}{N} \right)- \frac{\pi^2}{36} \left(\frac{k}{N}\right)^2 + \ldots \right]\,.
\ee
This difference is related to the fact that one ambient CFT is strongly coupled SYM while the other one is a system of free fermions. It would be interesting to have a better understanding of these behaviors.

Lastly, we mention that there is a discontinuity at precisely zero temperature, since any nonzero value is equivalent to a large temperature phase. The zero-temperature entropy can be computed using the fact that the configuration is supersymmetric and hence one just needs to count the number of
supersymmetric ground states with the appropriate defect fermion number.  As in \cite{KKYone}, the answer  is:
\begin{equation}
\mc S_\text{imp}[T = 0]  = \log \left[\frac{N!}{k!(N-k)!}\right]\,.
\end{equation}
This result is exact, and is related to the fact that the expectation value of the straight Wilson line is protected by supersymmetry from quantum corrections.

\section{Spectrum of defect operators and holographic excitations}\label{sec:excitations}

The $\mc N=4$ theory with a defect has gauge-invariant operators localized on the defect, which contain the defect fermion $\chi$. The dimensions of these operators determine important properties of the Kondo model, such as the stability of the fixed point and defect thermodynamics. This section is devoted to the analysis of defect operators and their spectrum. After reviewing their basic properties using the $OSp(4^*|4)$ algebra, we present an explicit calculation of the anomalous dimensions in terms of KK fluctuations of the D5 branes in the gravity dual. These results will be used in \S \ref{sec:thermo} to study the stability and thermodynamic aspects of the Kondo fixed point. The basic properties of the KK spectrum have recently been conjectured in \cite{UMich} using group theory analysis. Here we build upon their results, performing an explicit calculation of the D5 fluctuations. This will also show what representations of the algebra are physically realized.

\subsection{Defect operators}\label{subsec:operators}

Starting from the action (\ref{eq:fullS}) and (\ref{eq:Sdefect1}), at large $N$ the defect operators of lowest dimensions are single traces of $SU(N)$ indices and contain one pair of defect fermions,
\be
(\mc O^{(n)})_I^{\;J} = \sum_{i,j} \bar \chi_j^J \left(\phi^{a_1} \ldots \phi^{a_n} \right)_i^{\;j} \chi_I^i 
\ee
where $\phi^a$ are the $\mc N=4$ scalars. These operators transform in the adjoint of the flavor $U(M)$ group. Focusing on the spectrum at large $N$, the operators of lowest dimension arise from small multiplets of the $OSp(4^*|4)$ superconformal algebra preserved by the defect. These correspond to chiral primaries, whose dimensions are protected from quantum corrections. The even subgroup is $SL(2, \mathbb R) \times SO(3) \times SO(5)$, and states are classified by the corresponding quantum numbers $(h,j; m_1, m_2)$, where $h$ is the $SL(2, \mathbb R)$ dimension, $j$ is the $SO(3)$ spin, and $(m_1, m_2)$ are the $SO(5)$ Dynkin labels of the state. Lowest weight representations for this supergroup were analyzed in \cite{OSp}.

The operator of lowest dimension arises from an ultrashort supermultiplet of $OSp(4^*|4)$. The full supermultiplet is
\be\label{eq:ultrashort}
(1,0;0,1) \oplus (3/2,1/2; 1,0) \oplus (2,1;0,0)\,.
\ee
The first state has dimension 1, is an $SO(3)$ singlet and transforms as a vector of $SO(5)$. The next state is a spacetime fermion, and transforms in the spinor representations of $SO(3)$ and $SO(5)$. The last state, of dimension 2, is an $SO(3)$ triplet and $SO(5)$ singlet. For our purposes, we will also need short multiplets with lowest weight state of the form $(f,0)$ under $SL(2, \mathbb R) \times SO(3)$; the multiplet content for $f \ge 2$ is
\bea\label{eq:short}
&&(f,0;0,f) \oplus (f+1/2,1/2;1,f-1) \oplus (f+1,1;0,f-1) \oplus \nonumber\\
&& (f+1,0;2,f-2) \oplus (f+3/2, 1/2;1,f-2) \oplus (f+2,0;0,f-2)\,.
\eea

\subsection{D5 fluctuations and KK reduction}\label{subsec:fluct}

Having understood the multiplet structure of the superconformal algebra, the next step is to find which of these operators are physically realized. These defect operators are dual to fluctuations of the D5 worldvolume fields in the $AdS_5 \times S^5$ background. So our strategy will be to calculate the spectrum of D5 fields explicitly, and use these results to determine the anomalous dimensions of defect operators via
\begin{eqnarray}
h^{scalar}&=& \frac{d}{2} + \frac{1}{2}\sqrt{d^2 + 4m^2}\nonumber\\
h^{vector}&=& \frac{d}{2} + \frac{1}{2}\sqrt{(d-2)^2 + 4m^2}\nonumber\\
h^{spinor}&=& \frac{d}{2} + |m|\,,
\end{eqnarray}
with $d=1$, corresponding to an effective theory on $AdS_2$. Our analysis will focus on bosonic fluctuations, since their fermionic parterns may be obtained using the superconformal generators.

The spectrum of fluctuations is obtained by linearizing the D5 action (\ref{eq:DBI}) around the embedding (\ref{eq:embeddingD5}). The worldvolume metric is (\ref{eq:g5}), and there is nonzero worldvolume flux (\ref{eq:wvflux}). The relevant fields for us are the 6d gauge fields $\t A_\mu$, and the fluctuations $\delta u$ and $\delta \theta$ around (\ref{eq:embeddingD5}). The $AdS_2$ gauge field and $\delta \theta$  were studied by \cite{Paredes}. The full analysis is presented in the Appendix, and here we summarize the main results.

We denote the $AdS_2$ and $S^4$ metrics by
\be
ds_{AdS_2}^2 = \frac{-dt^2 + dr^2}{r^2}\;\;,\;\;d\Omega_4^2 = g^{(S^4)}_{ij} d\xi^i d\xi^j \,.
\ee
The 6d worldvolume coordinates are chosen to be $\xi^0 = t$, $\xi^1 = r$ and $\xi^i$ (the angles on $S^4$). The relevant fluctuations are
\be
\delta(ds^2) = \delta u^2 ds_{AdS_2}^2 + (\sin 2\theta_k\,\delta \theta+ \cos 2 \theta_k \, \delta \theta^2)d\Omega_4^2+ (d \delta u)^2+ (d \delta \theta)^2\;\;,\;\;F = \cos \theta_k \frac{dt \wedge dr}{r^2} + f\,,
\ee
where $f$ is a 2-form along 6d space. At this stage, the fields $\delta u$, $\delta \theta$ and $f$ are functions of the 6d coordinates $\xi^\mu$.

The DBI action up to quadratic order is found to be (see (\ref{eq:AfinalS}))
\bea\label{eq:finalS}
S^{(2)} &=& T_5 \int d^6 \xi \sqrt{g^{(S^4)}} \left\lbrace \frac{\sin^3 \theta_k}{r^2} \frac{1}{2}  \left[r^2 (\partial_t \phi_a)^2 - (\partial_r \phi_a)^2 - \partial^i \phi_a \partial_i \phi_a - 2 \delta u^2 + 4 \delta \theta^2 \right] \right. \nonumber\\
&+& \left.\frac{\sin \theta_k}{2} \left(r^2 f_{rt}^2 - \frac{1}{r^2} f_{ij}^2 - f_{ti} f^i_{\;t}+ f_{ri}f^i_{\;r} \right) - 4 \sin^2 \theta_k\,\delta \theta\,f_{rt} \right\rbrace\,.
\eea
We point out that the effective 6d metric that appears in this action is modified by the presence of worldvolume flux. Indeed, defining
\be
\hat g_{\mu\nu} dx^\mu dx^\nu \equiv (g_{\mu\nu}-F_{\mu \rho} F^\rho_{\;\nu}) dx^\mu dx^\nu= \sin^2 \theta_k \left( ds_{AdS_2}^2 + d\Omega_2^2\right)
\ee
(\ref{eq:finalS}) becomes
\bea
S^{(2)}&=& \frac{T_5}{\sin \theta_k} \int d^6 \xi \,\sqrt{- \hat g} \left\lbrace - \frac{1}{2} \hat g^{\mu\nu} \partial_\mu \phi_a \partial_\nu \phi_a - \frac{1}{2} \hat g^{\mu\nu} \hat g^{\rho \sigma} f_{\mu \rho} f_{\nu \sigma}+ \frac{1}{\sin^2 \theta_k}(-\delta u^2 + 2 \delta \theta^2)\right \rbrace \nonumber\\
&+& T_5 \int d^6 \xi (-4 \sin^2 \theta_k \, \delta \theta\,f_{rt})\,.
\eea
The modification of the worldvolume metric by flux is familiar in noncommutative field theories obtained from B-fields, and is expected on general grounds \cite{Martucci:2005rb}. By $SO(3)$ symmetry, $\delta u$ does not mix with the other modes; the mixed term between $\delta \theta$ and $f_{rt}$ agrees with the calculation in \cite{Paredes}. We also keep the other $f_{i \mu}$ and $f_{ij}$ components of the gauge field and we will explain shortly how to fix an adequate gauge. 

The next step is to perform the dimensional reduction, expanding the fluctuations in KK modes of $S^4$. The 6d scalars $\delta u$, $\delta \theta$ and $f_{rt}$ are expanded in terms of scalar spherical harmonics,
\be
\nabla_{S^4}^2 Y^l(y) =- l (l+3) Y^l(y)\,.
\ee
These modes transform in the $l$-th symmetric traceless representation of $SO(5)$ (denoted by $(0,l)$ in terms of the Dynkin labels). On the other hand, the gauge field components $a_i$ along $S^4$ decouple from the 2d gauge field by choosing the gauge $\nabla^i a_i=0$. These modes are expanded in terms of transverse vector spherical harmonics,
\be\label{eq:eigenv}
(\nabla_{S^4}^2-3) Y^l_i(y) = - l(l+1) Y_i^l(y)\,,
\ee
which transform as the $(2,l-2)$ of $SO(5)$. Details of the dimensional reduction may be found in appendix \ref{app:dimred}.

\subsection{Holographic dictionary for defect operators}\label{subsec:dictionary}

Let us summarize our results on the D5 fluctuations, and obtain the spectrum of defect operators. The fluctuations may be grouped in three sectors.
\vskip 2mm
$\bullet$ The $(\delta \theta, f_{rt})$ sector: these modes are mixed and, as we find in the Appendix and in \cite{Paredes}, their spectrum is 
\be\label{eq:deltasector}
m_l^2 = \left \lbrace 
\begin{matrix}
l(l-1)\qquad \;,\;\;l=1,2,\ldots \\
(l+3)(l+4)\;\;,\;\;l=0,1,2,\ldots 
\end{matrix}
 \right.\;\;\Rightarrow\;\;h_l =
 \left \lbrace 
\begin{matrix}
l\\
l+4
\end{matrix}
 \right.
\ee
Their quantum numbers are $(l+4,0;0,l)$ for $l \ge 0$, realizing the last state in (\ref{eq:short}), and $(l,0;0,l)$ for $l \ge 1$, corresponding to the first state in (\ref{eq:short}). In what follows, these two KK towers are denoted by $(\delta \theta, f_{rt})^{(1)}_l$ and $(\delta \theta, f_{rt})^{(2)}_l$ respectively.

\vskip 2mm
$\bullet$ $SO(3)$ triplet from D5 motion along $du^2 + u^2 d\Omega_2^2$. The $l$-th KK mode has mass
\be\label{eq:tripletsummary}
m_l^2 = 2+ l(l+3)\;\;,\;\;l \ge 0\,,
\ee
which is dual to an operator of dimension $h=l+2$. These modes realize operators with quantum numbers $(l+2,1;0,l)$ under $SL(2, \mathbb R) \times SO(3) \times SO(5) \subset OSp(4^*|4)$. This corresponds to the third state in the supermultiplet (\ref{eq:short}), after identifying $f \equiv l-1$. These modes will be denoted by $\delta u_l$.

\vskip 2mm

$\bullet$ The KK modes from the internal gauge field $a_i$ give rise to $SO(3)$ singlets with masses (see (\ref{eq:eigenv}))
\be\label{eq:internalgauge}
m_l^2 = l(l+1)\;\;,\;\;l \ge 2\,.
\ee
The dual operators have quantum numbers $(l+1,0;2,l-2)$, corresponding to the fourth state in (\ref{eq:short}). Below, we refer to these modes as $(a_i)_l$.

\vskip 1mm

Let us now establish the holographic dictionary for (chiral primary) defect operators. First, our analysis reveals that the lowest mode ($(\delta \theta, f_{rt})^{(1)}_{l=1}$ in the notation above) is dual to a marginal operator that transforms as an $SO(5)$ vector. This is consistent with the ultrashort multiplet (\ref{eq:ultrashort}) of the superconformal algebra. Geometrically, this mode corresponds to the D5 scalar field that rotates the embedding $SO(5) \subset SO(6)$. In the defect interaction $L \supset \bar \chi n^a \phi_a \chi$, this scalar field gives rise to a fluctuation $\delta n^a$ that is orthogonal to $n^a$. Since an $AdS$ field couples linearly to its dual operator, from this defect interaction we deduce that the operator dual to the massless scalar is $\bar \chi \phi_{\perp} \chi$, where $n^a \phi_{\perp\,,\,a}=0$.

At first sight, it may be a bit surprising that the only marginal operator that we find is an $SO(5)$ vector and not a singlet. Indeed, naively we would expect the Kondo interaction $\bar \chi (A_0 + n^a \phi_a) \chi$, which is an $SO(5)$ singlet, to give rise to a marginal deformation. However, the supersymmetry algebra fixes the coefficient of the Kondo coupling in terms of the ambient gauge coupling $g_{YM}$. Thus, turning on $\bar \chi (A_0 + n^a \phi_a) \chi$ without changing $g_{YM}$ breaks supersymmetry and the dimension of the operator is not protected. Our results indicate that at large $N$ this dimension diverges.

The holographic dictionary may be found starting from the lowest weight mode $(\delta \theta, f_{rt})^{(1)}_l$, which we just argued is dual to $\bar \chi \phi_{\perp}^l \chi$, and then recognizing that the rest of the modes fill in a short multiplet of $OSp(4^*|4)$ --see (\ref{eq:ultrashort}) and (\ref{eq:short}). We summarize the dictionary for defect operators in (\ref{table:dictionary}).\footnote{Here $\alpha=1,2,3$ are the space coordinates of the four-dimensional spacetime, while the index $i$ labels the $S^4$ coordinates (where the D5s are wrapped). Also, $\phi_\perp^a$ denotes the $\mc N=4$ scalar fields that transform as an $SO(5)$ vector, while the remaining scalar is $n^a \phi_a$. $( \phi_\perp^{(a_1} \ldots \phi_\perp^{a_l)} )$ is a symmetric traceless product, and the symmetrization in $(a_i)_l$ corresponds to the vector harmonics of $S^4$, explained in \S \ref{app:dimred}.} Only the bosonic fields are shown, and the operator content is schematic. The first two states realize the ultrashort multiplet, while the next four entries, defined only for $l \ge 2$, yield (\ref{eq:short}).
\begin{center}
\be\label{table:dictionary}
\begin{tabular}{c|c|c}
D5 field & defect operator & $SL(2, \mathbb R) \times SO(3) \times SO(5)$  \\
\hline
&&\\[-12pt]
$(\delta \theta, f_{rt})^{(1)}_{l=1}$&$ \mc O \equiv \bar \chi \phi_\perp \chi$ & $(1,0;0,1)$  \\
$\delta u_{l=0}$&$ Q^2 \mc O \sim \bar \chi (n^a D_\alpha \phi_a) \chi$ & $(2,1;0,0)$ \\
\hline
&&\\[-12pt]
$(\delta \theta, f_{rt})^{(1)}_l$&$ \mc O^{(l)} \equiv \bar \chi( \phi_\perp^{(a_1} \ldots \phi_\perp^{a_l)} )\chi$ & $(l,0;0,l)$  \\
$\delta u_{l-1}$&$ Q^2 \mc O^{(l)} \sim \bar \chi (n^a D_\alpha \phi_a \phi_\perp^{(a_1} \ldots \phi_\perp^{a_{l-1})} ) \chi$ & $(l+1,1;0,l-1)$ \\
$(a_i)_l$&$  Q^2 \mc O^{(l)} \sim \bar \chi(\Gamma_i n^a \phi_a\,\phi_\perp^{[a_1} \phi_\perp^{(a_2]} \ldots \phi_\perp^{a_l)} )\chi$ & $(l+1,0;2,l-2)$  \\
 $(\delta \theta, f_{rt})^{(2)}_{l}$&$ Q^4\mc O^{(l)} \sim \bar \chi((n^a D_\alpha \phi_a)^2 \phi_\perp^{(a_1} \ldots \phi_\perp^{a_{l-2})} )\chi$ & $(l+2,0;0,l-2)$
\end{tabular}
\ee
\end{center}

This concludes our analysis of chiral primary defect operators, that appear in the large $N$ spectrum. It is important to point out that, while the spectrum of fluctuations was derived in the probe approximation, the results are in fact exact. This is because the dimensions of chiral primaries are protected from quantum corrections. Next we will use these results to study the stability of the Kondo fixed point, and determine the thermodynamic properties induced by irrelevant deformations.

\section{Stability and defect thermodynamics}\label{sec:thermo}

So far we have studied the Kondo fixed point that arises from placing a fermionic impurity in $\mc N=4$ SYM, in a way that preserves half of the supersymmetries of the ambient theory. We determined the impurity entropy of the system and the spectrum of defect operators. The most interesting properties of the defect model are controlled by the lowest dimension operators allowed by symmetry that provide a correction to the scaling limit. In this section we use the results on the spectrum of operators to explore the stability of the fixed point and its basic thermodynamic observables.

\subsection{Stability of the Kondo fixed point}\label{subsec:stability}

The first point to address is whether the Kondo fixed point that we have found is stable; this would be the case if there are no relevant operators allowed by symmetries. It is useful to first review the situation in the real Kondo model (\ref{eq:H1}), based on free electrons interacting with a spin. In the original one channel $SU(2)$ spin model, the strongly coupled point $J \to \infty$ is stable, since by symmetry no relevant operators are allowed. In contrast, in the multichannel case this infinite coupling point is unstable (the ground state degeneracy at strong coupling is larger than for $J=0$). The model flows instead to an intermediate fixed point that is not described by Fermi-liquid theory.

Let us ask the same question in our case. Given the results (\ref{table:dictionary}) on the spectrum of defect operators, the operator $\mc O= \bar \chi \phi_\perp \chi$ of lowest dimension is marginal but it transforms as a vector under $SO(5)$. The lowest dimension singlet operator has then dimension 2, corresponding to $\mc O^2$. This is irrelevant in a $0+1$-dimensional theory. Thus, there are no relevant or marginal operators allowed by $SO(3) \times SO(5)$ and the Kondo fixed point is stable at large 't Hooft coupling. At small $\lambda$, operators that are not chiral primaries may also have small dimension and should be taken into account.

Notice that at large $N$ and $\lambda$, the stability of the fixed point is already guaranteed by the large $OSp(4^*|4)$ superconformal algebra. Indeed, in this limit the operators of lowest dimension arise from small representations, but this algebra does not have short multiplets with states of dimension $h \le 1$ that are singlets under $SO(3) \times SO(5)$. This should be contrasted with known CFTs that contain light scalars and thus can have marginal or relevant deformations. The restrictions from the global symmetries are relaxed in systems with impurities separated along the spatial direction. This breaks $SO(3)$, so operators with nonzero $SO(3)$ spin are now allowed in studying the stability of the impurity fixed point. In this case the lowest allowed dimension is two, from the second state in (\ref{table:dictionary}). So we learn that the impurity fixed point in a system with broken $SO(3)$ is still stable.

\subsection{Impurity specific heat and magnetic susceptibility}\label{subsec:thermo}

Interesting observables in a defect model include quantities like the impurity free energy, specific heat and magnetic susceptibility.  
In the strict scaling limit, because the free energy of the defect will scale as $F_{\rm defect} \sim T$ (see (\ref{eq:FD5})), the impurity specific heat
\begin{equation}
C_{\rm defect} = -T {\partial^2   F_{\rm defect} \over \partial T^2}
\end{equation}
will vanish.  

The behavior of the defect susceptibility is more subtle and interesting. We define the susceptibility as coming from considering
a background ``magnetic field'' $B$ that couples to the unbroken $SO(5)$ R-symmetry current via 
\be\label{eq:magnetic}
S \supset \int d^4 x\, \mc A_\alpha J_R^\alpha
\ee 
(with $\mc A$ the vector potential for B). The total susceptibility is then
\begin{equation}\label{eq:chitotal}
\chi_{\rm total} \equiv \left. {\partial^2 F \over \partial B^2} \right|_{B=0}~,
\end{equation}
and the impurity susceptibility is the piece of $\chi_\text{total}$ that does not scale like a spatial volume, namely $\chi_\text{defect} = \chi_\text{total} - \chi_\text{ambient}$. With our definition (\ref{eq:magnetic}) of magnetic field, the susceptibility is given by integrating the two-point function of $SO(5)$ R-currents. In the original overscreened Kondo model the defect susceptibility vanishes at the fixed point, but this need not be the case when the ambient theory has interactions. For instance, in the models of \cite{sachdevthree}, $\chi_\text{defect}=C/T$ with $C$ a universal (irrational) number. For us $\chi_\text{defect}$ vanishes trivially in the probe limit; however, results with circular Wilson loops in \S \ref{subsec:correlator} suggest that it is nonzero after taking into account the defect backreaction, with a coefficient proportional to (\ref{eq:Simpprobe}). Until we are ready to discuss the backreacted solution, we restrict the analysis of the susceptibility to the probe limit. In contrast, our results here for the specific heat are valid beyond the probe approximation.

In the overscreened Kondo model, the defect specific heat and impurity entropy vanish in the strict scaling limit.
It is because of this, as emphasized in e.g. \cite{Ludwig, Affleck}, that the crucial role in governing the defect thermodynamics is
played by the {\it leading~irrelevant~operator} that may be present in an RG flow to the critical point. In the overscreened model summarized in \S \ref{subsec:multichannel}, this operator is a descendant $\mc O_0 = \vec{J}_{-1} \cdot \vec{\phi}$ obtained by contracting the spin current with the primary operator in the adjoint of the spin group. The dimension of $\mc O_0$ is $h_0 = 1 + \frac{N}{N+K}$, which becomes $h_0=3/2$ for $K=N$. We stress again that for us $\chi_\text{defect}=0$ only in the probe limit --below we find a nonzero susceptibility in the backreacted case, even in the absence of irrelevant deformations.

We should therefore consider
the results for $ C_{\rm defect}$ and $\chi_{\rm defect}$ in the presence of such a perturbation. More precisely, one is generally interested in the leading operator consistent with the $SO(3) \times SO(5)$ global spin and flavor symmetries. Let us first discuss this for a general operator $\mc O_0$ that is a singlet under the global symmetries and has scaling dimension $h_0 \ge 2$ (recall that there are no relevant or marginal singlet operators). In the dual field theory, we imagine deforming the defect action so that
\begin{equation}
S_{\rm defect} \to S_{\rm defect} +  \int \,dt\, ( \lambda_0 {\cal O}_{0} + \text{h.c.})\,.
\end{equation}
Here, $\lambda_0$ is a parameter of mass dimension $1-h_0<0$; we can use it to define a scale analogous to the Kondo temperature in \cite{Ludwig,Affleck}, $T_K \sim \lambda_0^{-1/(h_0-1)}$. The defect operators considered below have vanishing one point function so, as in the overscreened Kondo problem, the corrections to the specific heat and susceptibility  are second order in $\lambda_0$. Now, the basic thermodynamic properties of the defect theory follow from dimensional analysis, recalling that the impurity specific heat is dimensionless and the susceptibility has dimensions of inverse length. At order $\lambda_0^2$, the contributions then read
\be\label{eq:C-chi-scaling}
C_{\rm defect} \sim \left({T \over T_K}\right)^{2(h_0-1)}\;,\;\;\chi_{\rm defect} \sim \left({T \over T_K}\right)^{2(h_0-1)} \frac{1}{T}\,.
\ee

Moving next to the specific form of the irrelevant perturbation, the analysis of KK fluctuations above shows that
the lowest operator allowed by symmetries is the double-trace
\begin{equation} 
 (\chi^\dagger \phi_\perp \chi) \cdot (\chi^\dagger \phi_\perp \chi)\,,
\end{equation}
with a sum over the $SO(5)$ indices of $\phi_\perp$. This operator has $h_0=2$, which curiously agrees with the normal Fermi liquid case (although in the Fermi liquid analysis the contribution was first order and not second order as here, so $C_\text{defect}$ and $\chi_\text{defect}$ scale differently with temperature). On the other hand, the overscreened model relevant for us has a leading operator of fractional dimension $h_0=3/2$. This would give rise to a linear scaling $C_\text{defect} \sim T$, as in the Fermi liquid case, although of course other properties of the system are sensitive to the non Fermi liquid nature of the fixed point.

Large $N$ correlators involving double trace operators are suppressed by additional powers of $N$ as compared to their single-trace counterparts. Thus, we may instead consider the leading single-trace singlet, which corresponds to the last state in (\ref{table:dictionary}). This operator is a descendant of $\mc O^{(2)}=\bar \chi(\phi_\perp^{(a_1} \phi_\perp^{a_2)}) \chi$ and has dimension $h_0=4$. It is also interesting to envision the possibility that the two irrelevant operators are simultaneously present in the approach to the fixed point. This would produce a transition in the impurity critical exponents, at a parametrically small temperature controlled by a power of $1/N$.

\section{Beyond the probe approximation: Kondo models and geometric transitions}\label{sec:backreacted}

Our analysis of the supersymmetric Kondo model has been restricted so far to the probe limit for the D5 branes. In the field theory side this corresponds to a quenched approximation that ignores quantum effects from the defect fields, and in the gravity side it amounts to treating the D5 branes as probes in the D3 geometry. Moreoever, we found that even starting from this limit it is possible to derive results that are valid beyond the probe approximation --these include the spectrum of defect operators and the thermodynamic properties presented in \S \ref{sec:thermo}.

Nevertheless, it is important to go beyond the probe approximation and understand how the insertion of the impurity deforms the ambient gauge theory and its correlators. These effects become important for $g_{YM}^2 M \gtrsim 1$. What type of field theory do we expect after taking into account the backreaction from the impurity? In the boundary CFT approach, the impurity disappears completely and is replaced by appropriate boundary conditions on the ambient fields. In this section we study the backreacted solution for the supersymmetric model, and find a realization of this effect in terms of a geometric transition in the gravity description. The gravity solution can be trusted when all the cycles are large, which in particular requires $g_{YM}^2 M \gg 1$.

\subsection{Basic properties of the backreacted solution}\label{subsec:basic}

Fortunately, the full solution including backreaction from the Wilson line defect is already known. The gravitational description of antisymmetric Wilson loops was first considered in \cite{Ybubble, Lunin}, and, impressively, the complete gravity solution was found by \cite{DEG}. It is described by a ``bubbling geometry'' of the type introduced in \cite{LLM}. Backreaction of the impurity can also be calculated in terms of a matrix model for the zero mode of $A_0+n_a \phi^a$ (when placing the theory on $S^4$).\footnote{As recognized by \cite{Semenoff,Gross}, matrix models provide powerful techniques to compute exactly certain quantities in $\mc N=4$ SYM. This connection was recently proved by \cite{Pestun}; the matrix model for antisymmetric Wilson loops was developed in \cite{Hartnoll:2006hr,Yamaguchi,Okuda:2008px,Gomis:2008qa}} Here we will summarize the gravity solution of \cite{DEG}, reserving some comments on the matrix model approach to the last part of the section.

Physically, taking into account backreaction replaces the D5 branes by 3-form field strengths $F_3$ and $H_3$, and the gravity solution does not include explicit source branes. This requires the appearance of new topologically nontrivial cycles (``bubbles'') to support the fluxes. This is known as a \textit{geometric transition}. Recalling the description of the $AdS_5 \times S^5$ limit in terms of (\ref{eq:AdSfiber1}) and since the full solution should still have $SL(2, \mathbb R) \times SO(3) \times SO(5)$ symmetry, the most general metric is of the form
\be\label{eq:generalmetric}
\frac{ds^2}{R^2}= f_1^2 ds_{AdS_2}^2 + f_2^2 d\Omega_2^2 + f_4^2 d\Omega^4 + d\Sigma^2\,,
\ee 
where $\Sigma$ is a Riemann surface with boundary (see below), and the radii $f_i$ are functions on $\Sigma$. Locally the space is of the form $AdS_2 \times S^2 \times S^4 \times \Sigma$, and the factors $AdS_2 \times S^2 \times S^4$ are fibered over $\Sigma$. In the $AdS_5 \times S^5$ case, which we will revisit shortly, the Riemann surface is given by $(u,\,\theta)$, and $d\Sigma^2 = du^2 + d \theta^2$.

In \cite{DEG} it was found that the full solution is determined in terms of two real harmonic functions $h_1$ and $h_2$ on $\Sigma$, given by the combinations
\be
h_1^2 = \frac{1}{4} e^{-\phi} f_1^2 f_4^2\;,\;h_2^2 =\frac{1}{4} e^{\phi} f_2^2 f_4^2\,,
\ee
where $\phi$ is the dilaton. The $f_i$ and the dilaton are nonvanishing inside $\Sigma$, and the solution has a single asymptotic $AdS_5 \times S^5$ (corresponding to moving far away from the impurity). The boundary $\partial \Sigma$ is defined by the locus where either $S^2$ or $S^4$ shrink to zero. Since $f_2$ or $f_4$ vanish on the boundary, $h_2=0$ everywhere on $\partial \Sigma$. Furthermore, in regular solutions $f_1$ is nonvanishing on $\partial \Sigma$; so $h_1=0$ on segments on $\partial \Sigma$ only when $S^4$ shrinks to zero. Given this structure, a topologically nontrivial 3-cycle is obtained by fibering $S^2$ over a one-cycle in $\Sigma$ that connects two boundary segments where $f_2=0$. Similarly, a homology 5-sphere can be constructed by fibering $S^4$ over a cycle between two boundary segments where $f_4=0$. These are the cycles that support nonzero fluxes and represent the backreaction of the branes on the geometry. A concrete solution is studied in detail in \S \ref{subsec:general}, and the topology is illustrated in Fig. \ref{fig:transition}.

Let us choose conformally flat coordinates on the Riemann surface,
\be
d\Sigma^2 = 4 \rho^2 dv d \bar v\,.
\ee
It is also convenient to introduce the combinations
\bea\label{Wis}
W &=& \partial h_1 \bar\partial h_2 + \bar \partial h_1 \partial h_2 \nonumber\\
V &=& \partial h_1 \bar\partial h_2 - \bar\partial h_1 \partial h_2 \nonumber\\
N_1 &=& 2 h_1 h_2 \vert \partial h_1\vert^2 - h_1^2 W \nonumber\\
N_2 &=& 2 h_1 h_2 \vert \partial h_2\vert^2 - h_2^2 W~.
\eea
In terms of these quantities, the metric functions $f_i$ and $\rho$ are given by~\cite{DEG}
\bea\label{eq:metricf}
f_1 &=& \left( - 4 \sqrt{-N_2 \over N_1} h_1^4 {W \over N_1}\right)^{1/4}\;,\;f_2 = \left(4 \sqrt{-N_1 \over N_2} h_2^4 {W\over N_2}\right)^{1/4} \nonumber\\
f_4 &=& \left( 4 \sqrt{-N_2 \over N_1} {N_2 \over W}\right)^{1/4}\;,\;\rho = \left(- {W^2 N_1 N_2\over h_1^4 h_2^4} \right)^{1/8}\,.
\eea
The dilaton $g_s^2 = e^{2\phi}$ is determined by
\begin{equation}
e^{2\phi} = -{N_2 \over N_1} > 0~.
\end{equation}

In what follows we first formulate $AdS_5 \times S^5$ in this language. Then explain how to choose $h_1$ and $h_2$ and consider and study the simplest case of one new 3-cycle.

\subsubsection{$AdS_5 \times S^5$ warm-up}

In order to develop intuition about the boundary conditions that arise on $\partial \Sigma$ and the form that $h_1$ and $h_2$ can take, it is instructive to first consider the $AdS_5 \times S^5$ case. In the general solution, this may be seen as the asymptotic description far away from the D5 insertions.

The $AdS_5 \times S^5$ metric was written in the form (\ref{eq:generalmetric}) in \S \ref{subsec:probegravity}. In this case the Riemann surface is a semi-cylinder $0\le u< \infty$, $-\pi/2\le \theta \le \pi/2$. Note here that at each point on the boundary of the $\Sigma$ cylinder, the volume of either the $S^2$ or the $S^4$ vanishes. Along the locus $\theta = \pm {\pi \over 2}$ it is the $S^4$ that vanishes, while along the locus $u=0$ it is the $S^2$.

To make contact with the more general case, it is useful to define a new complex coordinate $v$ via
\begin{equation}
\label{vis}
v = - i ~{\rm sinh}(u + i\theta)~.
\end{equation}
The full boundary of the cylinder is now mapped to the line ${\rm Im}\,v = 0$.  For ${\rm Re} \, v < -1$ and ${\rm Re}\,v > +1$, $f_4$ vanishes;
for $-1 < {\rm Re}\,v < +1$, $f_2 =  0$.  The ``central line'' $\theta=0$ in the $(u, \theta)$ cylinder is mapped to the negative imaginary axis, and the whole $(u, \theta)$
cylinder is mapped to the lower half-plane in the $v$ coordinate.
The structure of the boundary is therefore a line along which loci with vanishing $S^4$ are separated by a locus with vanishing $S^2$.  Drawing a 
curve into the interior connecting the distinct loci where the $S^4$ vanishes, one sweeps out a topologically non-trivial $S^5$.

In order to verify
that one has proper behavior at the asymptotics in more non-trivial cases with $D5$ branes, it is useful to map the asymptotic $v \to \infty$ region to a finite point. This can be accomplished by a final change of variables, defining:
\begin{equation}
\label{uis}
w - w_0 = -{1 \over {1+v}}~.
\end{equation}
Here, $w_0$ is an arbitrary finite point, and $w=w_0$ is where the $AdS_5 \times S^5$ asymptotic region is mapped.  So in the more general cases with D5 branes, even after backreaction, one will impose the condition that the solution asymptotes to $AdS_5 \times S^5$ as $w \to w_0$.
We will be more concrete about what this means, after developing a bit more technology.

Note that in the new coordinates, if we choose $w_0$ to be ${\it real}$, we find that the region on the real axis where $h_1 \neq 0$ in the $v$-variable, is mapped
to the region $(-\infty, e_1)$ on the real axis for the $w$-variable, with $e_1=w_0-1/2$.
In fact, the Riemann surface $\Sigma$ can be viewed as the surface defined by the equation
\begin{equation}
\label{genuszero}
s(w)^2 = w - e_1~.
\end{equation}
We will henceforth always assume, in all cases, that $w_0$ is chosen to be real.

In this notation, $AdS_5 \times S^5$ corresponds to
\begin{equation}
h_1 = 2\,{\rm cosh}\, u\, {\rm cos} \,\theta ,~~h_2 =2\, {\rm sinh}\, u \, {\rm cos} \,\theta~.
\end{equation}
In terms of the $v$ coordinates, this becomes
\begin{equation}
h_1 = \sqrt{1-v^2} + c.c.,~~h_2 = i(v- \bar v) ~.
\end{equation}
Going through the final change of variables (\ref{uis}) obtains
\begin{equation}
\label{htrivial}
h_1 = \sqrt{2} {i \over w-w_0} \sqrt{w - e_1} + c.c.,~~h_2 = i \left(-{1\over w-w_0} + {1\over \bar w - w_0} \right)~.
\end{equation}
Notice that the branch cuts are on the real axis, and that $h_1$ is precisely non-vanishing on the real axis when $w > e_1.$  

In the remainder of the paper, we will switch back and forth between the $v$ and $w$ variables, depending on which is more convenient for a 
given discussion.

\subsection{General formalism for backreacted solutions}\label{subsec:general}

This structure will
generalise to cases including D5 stacks.  The asymptotically $AdS_5 \times S^5$ nature of all of the relevant geometries means that the regions near $v = \pm \infty$ will
always have $f_4 = 0$; if there are $p$ intervening regions where $f_2 = 0$, then there will be $p$ topologically non-trivial $S^5$s in the geometry,
and $p-1$ topologically non-trivial $S^3$s.  
Not surprisingly, $p-1$ then maps to the number of distinct $D5$ stacks (wrapping different polar angles along the $S^5$) that we include, all sitting at the
same point in $\mathbb R^3$.  We will focus on the case $p=2$, so in the backreacted geometry, the ${\rm Im}\,v=0$ line will be split into 5 regions: 3 segments 
with vanishing $f_4$, and two interior regions with vanishing $f_2$.  The $F_3$ flux through the single non-trivial $S^3$ will encode the number of
D5s in the stack.

Let us now describe in detail the case of a single stack of $M$ D5 branes, namely when $\Sigma$ is of genus 1.
We see from (\ref{htrivial}) that in the simplest case (representing just an unperturbed $AdS_5 \times S^5$ geometry), one has
\begin{equation}
dh_2 = i {dw \over (w-w_0)^2}~.
\end{equation}
That is, $dh_2$ has a double-pole at $w=w_0$.  In fact, \cite{DEG} found that this should ${\it always}$ be the case for regular solutions that are asymptotically $AdS_5 \times S^5$.
So the only change from the backreaction of the D5 branes occurs in the harmonic function $h_1$.

\begin{figure}[htb]
\begin{center}     
\includegraphics[width=0.7\textwidth]{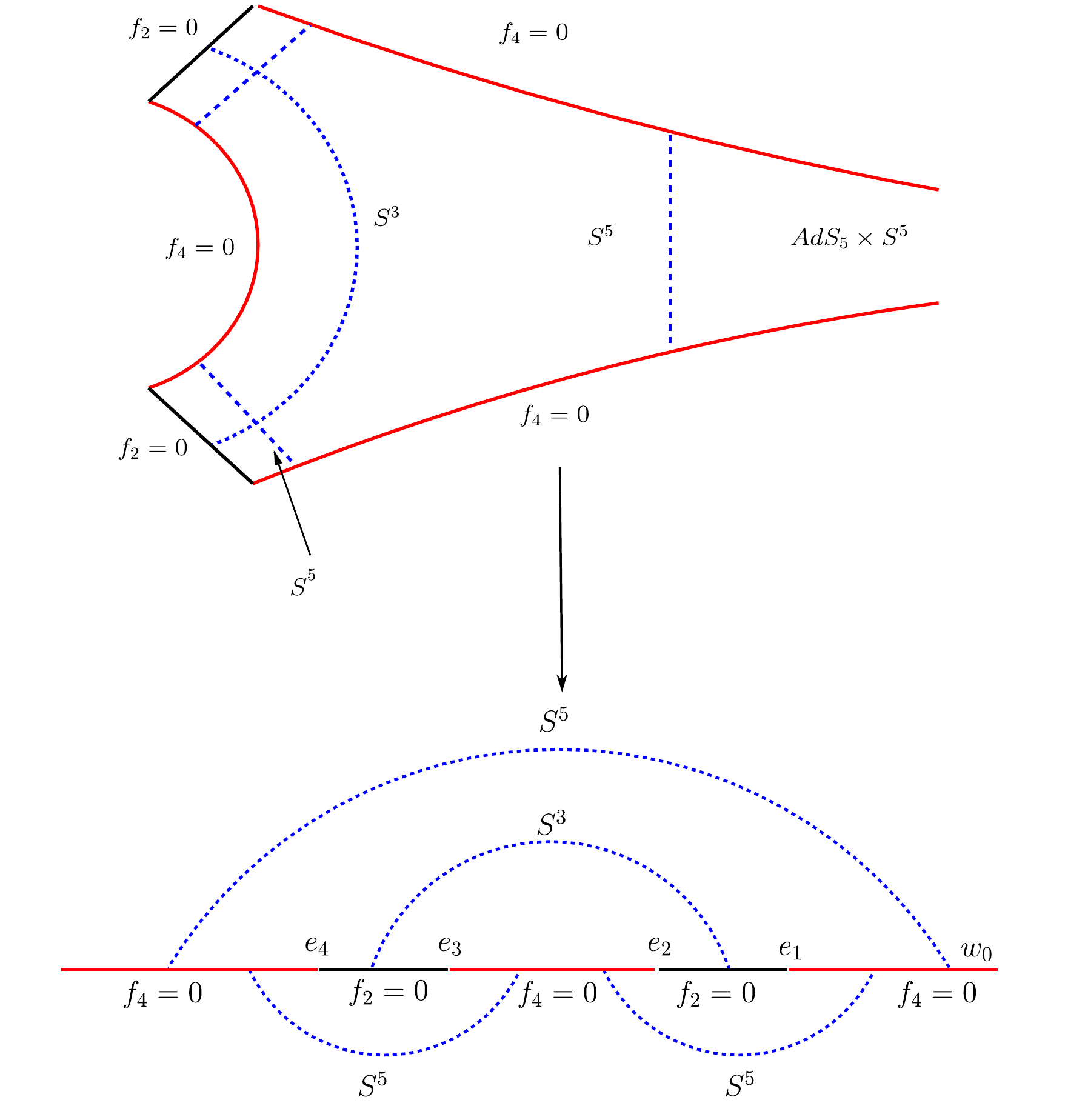}
\caption{\small{Backreacted solution in the presence of a single D5 stack. The upper figure shows how the original cylinder corresponding to $AdS_5 \times S^5$ is deformed by the introduction of D5 branes, which create a new boundary segment with $f_4=0$. The lower figure shows the Riemann surface with boundary after the conformal transformations described in the text. The topologically nontrivial 3- and 5-cycles are shown in dotted blue lines.}}
\label{fig:transition}
\end{center}
\end{figure}

Now, as in \S2, let us pick some angle $\theta$ on the $S^5$, and wrap M D5-branes there.
Along the real $w$ axis, we expect three segments where $h_1 = 0$, broken up by two finite regions where $h_1 \neq 0$.  If we call the real
coordinates where the transitions between $h_1 = 0$ and $h_1 \neq 0$ take place $e_{1,2,3,4}$, then without loss of generality we can
set $e_4 = \infty$. The Riemann surface with boundary and the homology 3- and 5-cycles are shown in Fig. \ref{fig:transition}.

The defining equation of the genus one Riemann surface (where $h_1$, $h_2$ are single-valued) is given by
\begin{equation}
\label{torus}
s^2(w) = (w-e_1) (w-e_2) (w-e_3)~.
\end{equation}
We can always redefine the $w$ variable so that $e_1 + e_2 + e_3 = 0$, so there are two real parameters in the $e_i$.  
These parameters will govern the integrated fluxes over the non-trivial $S^3$ and $S^5$s in the geometry; they are related to the number of
D5 branes inserted, $M$, and the ``shift from N'' in the amount of $F_5$ flux through the two $S^5$s, which will be given by $N-k$ and $k$, with $k$
determined in terms of the angle of the $S^4$ which the D5s wrap (see (\ref{eq:thetak})).

In the form (\ref{torus}), the surface is still non-compact.  We can make it into a compact Riemann surface by going to Weierstrass form.  In terms of the 
Weierstrass ${\cal P}$ function, we can define the $z$ coordinate by the equation
\begin{equation}
w = {\cal P}(z)~.
\end{equation}
Since the Weierstrass function identically satisfies
\begin{equation}
{\cal P}^\prime(z)^2 = 4 ~\prod_{i=1}^{3} ({\cal P}(z) - e_i)~,
\end{equation}
we see that $s(w)$ automatically satisfies $s(w) = \frac{1}{2} {\cal P}^\prime (z)$.

Let us define the images in the $z$-plane of the special points in the $w$-plane via
\begin{equation}
w_0 = {\cal P}(z_0),~~e_i = {\cal P}(\omega_i)~.
\end{equation}
Also, define the Weierstrass $\zeta$ and $\sigma$ functions via
\begin{equation}
\zeta^\prime(z) = -{\cal P}(z),~~\zeta(z) = {\sigma^{\prime}(z) \over \sigma(z)}~.
\end{equation}
Then the harmonic functions are given by~\cite{DEG}
\begin{equation}
\label{honetorus}
h_1 = 2i \left( \zeta(z-z_0) + \zeta(z+z_0) - 2{\zeta(\omega_3) \over \omega_3} z - c.c. \right)
\end{equation}
and
\begin{equation}
h_2 = {i \over {\cal P}^\prime(z_0)} \left( \zeta (z+z_0) - \zeta (z-z_0) - c.c.\right)~.
\end{equation}
This is the desired final expression for $h_1$ and $h_2$, which provides an explicit solution for the single D5-stack case. Generalizations to more stacks can be similarly worked out in terms of hyperelliptic Riemann surfaces and harmonic functions on them.

The number of D5-branes and the $\theta$ angle of the $S^4$ in $S^5$ which they wrap, should be indicated by the amounts of three-form and five-form flux
threading the various topologically non-trivial spheres in the geometry.  In terms of the holomorphic and antiholomorphic parts
\begin{equation}
\label{parts}
h_1 = {\cal A} + \overline{\cal A},~~h_2 = {\cal B} + \overline{\cal B}\,,
\end{equation}
the expressions governing the fluxes are as follows \cite{DEG,Okuda:2008px}:
\begin{equation}
\label{threeflux}
\int_{S^3} F_3~ = ~4\pi^2 \alpha^\prime M = 8\pi \int_{e_2}^{e_3} ~(i \partial {\cal A} + c.c.)~
\end{equation}
and
\begin{equation}
\label{fivefluxes}
\int_{S^5_I} F_5 = 4\pi^4 (\alpha^\prime)^2 N_I = 8\pi^2 \int_{e_{2I}}^{e_{2I-1}} ({\cal A}\partial {\cal B} - {\cal B} \partial {\cal A} + c.c.)~.
\end{equation}
In (\ref{fivefluxes}), the two different 5-spheres are labeled by $I=1,2$, and the integration contour should actually be taken just a bit below the
real line segment running from $e_{2I}$ to $e_{2I-1}$ - i.e., with small negative imaginary component.

\subsection{Matrix model approach}

We just described the solution very explicitly in terms of elliptic functions and geometric transitions where new topologically nontrivial cycles appear. At this stage, it is natural to ask how all these properties are understood in field theory language. The simplest answer is given in terms of a matrix model that keeps the zero mode of $A_0 + \phi$ when the theory is placed on $S^4$. Intuitively, the distribution of eigenvalues in the matrix model will describe the homology cycles, and the Riemann surface will correspond to the spectral parameter. Let us briefly explain this approach, following \cite{Yamaguchi, Okuda:2008px,Gomis:2008qa}.

The claim is as follows. Let $M$ in this section denote an $N \times N$ Hermitian matrix associated to the zero mode of $A_0 + n_a \phi^a$.\footnote{This is not to be confused with a 
number of D5-branes, which was denoted by $M$ in earlier sections.} 
The expectation value of a gauge-invariant
function $f(M)$ --i.e. $f(V M V^{-1}) = f(M)$ for a unitary matrix $V$-- is given by
\begin{equation}
\label{expec}
\langle f(M) \rangle \equiv {1\over Z} \int~dM~f(M)~{\rm exp}\left(-{2N \over \lambda} {\rm tr}[M^2]\right)~,
\end{equation}
where the partition function $Z$ is defined by
\begin{equation}
Z \equiv \int~dM~{\rm exp}\left( -{2N \over \lambda} {\rm tr}[M^2]\right)~.
\end{equation}
Here the measure $dM$ denotes $\prod dM_{ij}$ subject to the constraint of hermiticity.
Then, the claim is that the BPS Wilson loop (whose action is identical to that discussed in \S2.2), in the $R$ representation of the gauge group, can
be calculated in the matrix model by computing $\langle \tr_{R} [e^M]\rangle$.

The partition function reduces to an integral of the exponential multipled by a familiar
Vandermonde determinant encoding the eigenvalue repulsion of the eigenvalues $m_i$ of $M$:
\begin{equation}
Z \sim \int ~\prod_{i=1}^N dm_j ~\Pi_{i\leq i <  j \leq N} ~(m_i - m_j)^2 {\rm exp} \left(- {2N \over \lambda} \sum_{j=1}^N m_j^2 \right)~.
\end{equation}
As usual, the eigenvalue repulsion can be exponentiated into the ``action" in the matrix integral, yielding a logarithmic potential:
\begin{equation}
Z \sim \int ~\prod_{i=1}^N dm_j~{\rm exp} (-S(m))
\end{equation}
\begin{equation}
S(m) \equiv {2N \over \lambda} \sum_{j=1}^N m_j^2 - 2 \sum_{i<j} {\rm log} \vert m_i - m_j \vert~.
\end{equation}

Consider now the insertion of the Wilson loop in the $k$-th antisymmetric representation; after diagonalizing $M$, the trace becomes
\begin{equation}
\tr_{A_k}[e^M] = \sum_{1 \leq j_1 < j_2 .<.. < j_k \leq N} ~{\rm exp}(m_{j_1} + ... + m_{j_k})~.
\end{equation}
As a result, we can also absorb the operator insertion into a new effective action and do a purely exponential integral:
\begin{equation}
\langle \tr_{A_k}[e^M]\rangle \sim \left(\frac{N!}{(N-k)!~ k!}\right) \int~\prod_{j=1}^N ~dm_j~{\rm exp}(-S_k(M))
\end{equation}
where
\begin{equation}
S_k(M) = S(M) - \sum_{i=1}^k ~m_i~.
\end{equation}

Introducing the resolvent
\be
\omega(z) = \tr\,\frac{1}{z-M}\,,
\ee
one can show that the equation of motion for $m_i$ relates $\omega$ to the harmonic functions of the bubbling solution via \cite{Okuda:2008px,Gomis:2008qa}
\be
h_1 \propto 2 (z - \bar z) - (\omega - \bar \omega)\;,\;h_2 \propto z - \bar z
\ee
once the spectral parameter $z$ is identified with the Riemann surface coordinate $v$. In this way, the geometric transition can be seen directly in the field theory.\footnote{In fact, it is easy to evaluate the integral by saddle point in the probe approximation, and solve for $S_k$ \cite{Yamaguchi}. The result agrees with the probe calculation in the gravity solution.}

\subsection{Comments on correlation functions}\label{subsec:correlator}

Having described how to calculate quantum corrections from backreaction of the defect, the next step is to compute bulk correlation functions. The simplest correlators are one-point functions which, in the gravity side, follow from expanding the background solution around the asymptotic $AdS_5 \times S^5$ region, and extracting the leading corrections. Conformal invariance restricts the expectation value of a chiral primary $\mc O_\Delta$ of dimension $\Delta$ to be of the form
\be\label{eq:1pt}
\langle \mc O_\Delta(x) \rangle_W = \frac{\Xi_\Delta}{l^\Delta}
\ee
where the subindex `$W$' denotes the correlator in the presence of the Wilson line insertion, $l$ is the distance between the line and the operator position, and $\Xi_\Delta$ is a constant that depends on $\Delta$ and the representation of the Wilson loop. All the nontrivial information is thus in $\Xi_\Delta$, which can be calculated either using the gravity solution or in terms of the matrix model. One-point functions of chiral primaries in the presence of a Wilson loop in the antisymmetric representation were calculated in \cite{Gomis:2008qa}.

Higher correlators are harder to compute and, to our knowledge, have not been obtained. It would be quite useful to fill in this gap, as these correlators contain important information about the response of $\mc N=4$ SYM to a nonlocal observable such as the Wilson loop. Let us outline the first steps in this program, which are also necessary for calculating transport properties away from the probe limit. We should start by understanding correlation functions of the gauge invariant chiral primary of lowest dimension, which is
\be
\mc O_2 \equiv \frac{8 \pi^2}{\lambda} \,\tr \left(\phi^a \phi^b - \frac{1}{6}\delta^{ab} \phi^a \phi^a \right)
\ee
where here $a, b$ are $SO(6)$ indices, and the normalization is such that in the planar limit $\langle \mc O_2(x) \mc O_2(y) \rangle = 1/|x-y|^4$. This dimension 2 operator is very important, because its descendants include the R-symmetry current $(J_\mu)^a_{\phantom{c}b}$ and energy momentum tensor $T_{\mu\nu}$. Therefore, the correlation functions for $\mc O_2$ determine, via superconformal transformations, the basic properties of the correlators of $J_\mu$ and $T_{\mu \nu}$ \cite{Dolan:2001tt}. The one point function coefficient in (\ref{eq:1pt}) is given by \cite{Gomis:2008qa}
\be\label{eq:Xi2}
\Xi_2 = \frac{\lambda^{1/2}}{6\pi} \sin^3 \theta_k\,,
\ee
which is proportional to the $g$-function (see (\ref{eq:Simpprobe})).

Approximate correlation functions can be obtained if we consider a circular Wilson loop (instead of a line operator), and expand it in terms of local operators \cite{Okuyama:2006jc}
\be\label{eq:Wope}
W_\text{circle}(a) = \langle W_\text{circle} \rangle\,\sum_\Delta\,\xi_\Delta\,(2\pi a)^\Delta \mc O_\Delta(0)
\ee
with $a$ the radius of the circle. This is valid for insertions at points far away from the circle. In this way, correlation functions with a Wilson loop insertion are approximated by higher point functions in the unperturbed $\mc N=4$ theory. Since a comparison between (\ref{eq:Wope}) and (\ref{eq:1pt}) reveals that for chiral primaries $\xi_\Delta \propto \Xi_\Delta$, in this approximation the correlators are determined by the one point functions that were discussed before.

As an application, consider the correlator $\langle J_\mu(x) J_\nu(x') \rangle_W$ for $SO(6)$ R-currents in the presence of a circular Wilson loop. Expanding the Wilson loop as in (\ref{eq:Wope}), the lowest dimension contribution is proportional to $a^2\Xi_2 \,\langle J_\mu J_\nu \mc O_2 \rangle$. The $\mc N=4$ correlator that we are left with can be obtained by applying superconformal transformations to the 3-point function of $\mc O_2$. 
This suggests that also the current-current correlator in the presence of a Polyakov loop will be non-zero, with $\chi_\text{defect} \sim \Xi_2/T$ which, given (\ref{eq:Xi2}), is proportional to the irrational $g$-function. This is reminiscent of the behavior found in \cite{sachdevthree}. On this note we end our anaysis, and hope to come back to the determination of correlators and transport properties in a future work.
 
 \section{Discussion}

We have seen in this paper that the most basic physical questions about the maximally supersymmetric Kondo problem can be answered by standard holographic techniques.
The operator content of the theory is determined by harmonic analysis on the geometry wrapped by the probe branes which represent the impurity in the gravity dual,
and
the backreaction of the impurity on the bulk liquid is encoded in a geometric transition.  Several of the properties of our model show striking resemblance to 
results on multi-channel Kondo problems in the field theory literature, but there are important differences, stemming from the power of the superconformal algebra
which controls the IR fixed point.  For instance, although some defect thermodynamic properties are most analogous to those of overscreened multi-channel
Kondo models, the leading irrelevant operator at the fixed point (which governs some of the most interesting physics of the defect) is dimension two, as in the
original single-channel Kondo problem.  

It is certainly of interest to extend the results of this paper to systems with less symmetry.  Clear directions for future research include:

\medskip
\noindent
$\bullet$
While several features of the model seem analogous to those of multi-channel Kondo models studied in the condensed matter literature, some important
differences arise.  Notably, instead of obtaining non-Fermi liquid scaling of the leading irrelevant operator, we obtain
a dimension $2$ operator, precisely the same as in the normal Fermi-liquid case (nevertheless, the contribution is second order in the coupling).  It seems likely that this
is because the superconformal algebra heavily constrains the spectrum of operators, resulting in the leading irrelevant $\Delta = 2$ operator we described in
\S5; certainly, the itinerant sector here has little resemblance to a single-channel bulk Fermi liquid!   
In similar models of impurities interacting with less symmetric CFTs than ${\cal N}=4$ supersymmetric Yang-Mills, we anticipate that the results will
be less constrained and that the coincidence in the behavior of the leading irrelevant operator will disappear.

\medskip
\noindent
$\bullet$
We have studied here the interactions between the defect and the ambient theory at vanishing chemical potential.  A more realistic analogue of the Kondo problem
would be to consider the ${\cal N}=4$ theory at finite $\mu$ for the $SO(6)$ R-charge; even more similarity to the real problem could be obtained by
using the ideas in \cite{Sachdevnew} to model a bulk Fermi liquid.  Needless to say, finding the finite temperature solutions could also be of interest.

\medskip
\noindent
$\bullet$ 
It would be very useful to study correlators of $\mc N=4$ SYM in the presence of a Wilson loop. To our knowledge, this has not been done yet. Furthermore, with the backreacted gravity solution in hand, we can now ask how the defect modifies bulk transport.  For instance, we can study the bulk resistivity
by computing current-current correlators in the modified background holographically.  Such studies, either in this microscopic model or in ``bottom-up"
constructions of similar models, might allow one to find interesting generalizations of the original ``resistance minimum" that motivated the tremendous
theoretical interest in the Kondo model to begin with.

\medskip
\noindent
$\bullet$ Our results motivate further study of overscreened models with equal number $N=K$ of spins and channels. In this case the ambient electrons become $N \times N$ matrices, and we described powerful matrix model techniques to solve this system. Studying more general large $N$ matrix models for the Kondo problem could add a valuable tool to the techniques that already exist in the literature. Even at the level of the $\mc N=4$ SYM, the matrix model description should be generalized to include defect operator insertions.

\medskip
\noindent
$\bullet$
Finally, and most ambitiously, one would like to push for a much more complete understanding of ${\it lattices}$ of such defects, including their backreaction
on the bulk and their interactions with one another.  The results in \cite{KKYtwo,Jensenetal} demonstrate that the simplest supersymmetric lattice models can give
rise to interesting non-Fermi liquid physics in the semi-holographic approach \cite{semi}.  But the analysis in \cite{Jensenetal} also suggests that backreaction at some
power-law suppressed scale in $1/N$ will destabilize the non-Fermi liquid (as it destroys the $AdS_2$ geometries which are at the root of the non-Fermi liquid
behaviour).  Our analysis here sheds interesting light on those results.  A priori, one might have thought that inter-defect interactions would lead to multiple
defects ``destabilising" each other's Kondo fixed point.  In RG language, the lattice model breaks the $SO(3)$ symmetry (but not necessarily the $SO(5)$ symmetry)
enjoyed by the single-defect fixed point.  Hence, defect-defect interactions could a priori induce relevant $SO(3)$-breaking operators in the fixed point
Lagrangian, destabilizing the fixed point.  However, a surprising result of our analysis in \S4\ is that ${\it even ~with~SO(3)~breaking}$, there are no relevant
defect operators at the fixed point.  This is not inconsistent with the discussion in \cite{Jensenetal}; in fact, it is not inter-defect interactions, but defect interactions
with the bulk that must destabilize the ${\it bulk}$ phase in the supersymmetric Kondo lattice.  This is plausible because the ${\cal N}=4$ theory has a marginal
operator that respects all relevant symmetries; in the dual gravity, it maps to the dilaton, and gravitational backreaction can source the dilaton, driving one out of
the gravity regime entirely.  This does suggest, however, that if one can construct similar defect models where the ambient theory of the itinerant modes does
not have marginal operators respecting all of the relevant symmetries, one may be able to find lattice models which remain stable, with $AdS_2$
regions on the lattice of impurity branes, order by order in $1/N$.

\section*{Acknowledgments}
We are grateful to A. Adams, N. Bobev, E. D'Hoker, K. Jensen, A. Karch, A. Ludwig, R. Mahajan, J. McGreevy, M. Mulligan, J. Polchinski, S. Sachdev, G. Semenoff, S. Shenker, E. Silverstein and S. Yaida for useful conversations about related subjects.
SK is grateful to the International Institute of Physics in Natal, the ICTP in Trieste, and the Aspen Center for Physics for hospitality while this work was in progress.
SH and SK acknowledge the Kavli Institute for Theoretical Physics, and the many participants in the ``Holographic Duality and Condensed Matter Physics''
workshop, for providing a very stimulating and supportive environment as this work was completed. GT would like to thank the Simons Center for Geometry and Physics for hospitality while this work was being completed. SH is supported by an ARCS Foundation, Inc. Stanford Graduate Fellowship. This research is supported in part by the US DOE under contract DE-AC02-76SF00515. 

\appendix

\section{Spectrum of fluctuations of the D5s}

In this Appendix we calculate the spectrum of fluctuations for the 6d theory on the D5 branes, placed in the background of the color D3 branes. We will first compute the quadratic 6d action from (\ref{eq:DBI}), and in a second step perform the dimensional reduction on $S^4$. The relevant quantities are the background 6d metric (\ref{eq:g5}), the worldvolume flux (\ref{eq:FD5}), the 4-form flux (\ref{eq:C4}), and the embedding (\ref{eq:embeddingD5}).

The bosonic fields on the 6d worldvolume are a gauge field $\t A_\mu$ and the scalar fluctuations around the embedding (\ref{eq:embeddingD5}). These are $\theta = \theta_k + \delta \theta$, $u = \delta u$, and two angle fluctuations along the $S^2$. Near $u=0$ we have $du^2 + \sinh^2u\,d\Omega_2 \approx ds_{\mathbb R^3}^2$, so $\delta u$ and the two angle fluctuations combine into a 3-vector $\varphi_m$. These transform as a fundamental of $SO(3)$, which becomes the R-symmetry of the 6d theory. In the original 4d spacetime, $\varphi_m$ is simply the fluctuation in the 3d spatial position of the defect. For the purpose of computing the spectrum, it is enough to study $\delta u$, since the masses of the triplet are then the same based on the $SO(3)$ symmetry.

\subsection{Six-dimensional quadratic action}

Let's start by expanding the determinant term of the DBI action in the fluctuations. Define $M_{\mu\nu} \equiv G_{\mu\nu}+ F_{\mu\nu}$. Then the background value is
\be
\overline{M}_{\mu\nu}= \left(
\begin{matrix}
-\frac{1}{r^2} & \frac{\cos \theta_k}{r^2} & 0_{1 \times 4} \\
-\frac{\cos \theta_k}{r^2} &  \frac{1}{r^2} & 0_{1 \times 4} \\
0_{4\times 1} & 0_{4\times 1} & \sin^2 \theta_k g^{(S^4)}
\end{matrix}
\right)
\ee
where the zero entries are blocks of the indicated size. The expansion of the first term of the DBI action gives
\be\label{eq:expansion}
\sqrt{-\det\,(\overline M + \delta M)} = \sqrt{-\det\,\overline M} \left[1+ \frac{1}{2} \tr(\overline M^{-1} \delta M)+ \frac{1}{8} (\tr(\overline M^{-1} \delta M))^2 - \frac{1}{4} \tr\,(\overline M^{-1} \delta M)^2 \right]
\ee

The metric fluctuation reads
\be
\delta(ds^2) = \delta u^2 ds_{AdS_2}^2 + (\sin 2\theta_k\,\delta \theta+ \cos 2 \theta_k \, \delta \theta^2)d\Omega_4^2+ (d \delta u)^2+ (d \delta \theta)^2\,,
\ee
and for the field strength we write
\be
F = \cos \theta_k \frac{dt \wedge dr}{r^2} + f\,,
\ee
where $f$ is a 2-form along 6d space. Thus,
\be
\delta M_{\mu\nu}= (\sin 2 \theta_k \,\delta \theta+\cos 2 \theta_k \, \delta \theta^2) g^{(S^4)}_{\mu\nu} + \delta u^2\,g^{(AdS_2)}_{\mu\nu}+ \partial_\mu(\delta u) \partial_\nu(\delta u)+\partial_\mu(\delta \theta) \partial_\nu(\delta \theta)+f_{\mu\nu}\,.
\ee

Raising an index we obtain
\bea\label{eq:delM}
\overline M^{\mu \rho} \delta M_{\rho \nu} &=& \frac{1}{\sin^2 \theta_k}
\left( 
\begin{matrix}
r^2 \cos \theta_k\,f_{rt} & r^2 f_{rt} & -r^2 f_{ti}+ r^2 \cos \theta_k\,f_{ri} \nonumber\\
r^2 f_{rt} &r^2 \cos \theta_k\,f_{rt} & -r^2 \cos \theta_k\,f_{ti}+ r^2 f_{ri}\nonumber\\
f^j_{\;t} & f^j_{\;r} & f^j_{\;i} + 2 \cos \theta_k\,\sin \theta_k \,\delta \theta \,\delta^i_{\;j}
\end{matrix}
\right) +\nonumber\\
&+& \frac{\delta u^2}{\sin^2 \theta_k}  
\left(
\begin{matrix}
1 &r^2 \cos \theta_k  & 0_{1 \times 4} \\
 \cos \theta_k  &  1 & 0_{1 \times 4} \\
0_{4\times 1} & 0_{4\times 1} & 0_{4 \times 4}
\end{matrix}
\right) +
\frac{\cos\,2\theta_k}{\sin^2 \theta_k}   \delta \theta^2
\left(
\begin{matrix}
0 &0  & 0_{1 \times 4} \\
 0  & 0 & 0_{1 \times 4} \\
0_{4\times 1} & 0_{4\times 1} & \delta^i_{\;j}
\end{matrix}
\right)+\nonumber\\
&+&\frac{1}{\sin^2 \theta_k}
\left( 
\begin{matrix}
r^2 \left[-U_{tt}+ \cos \theta_k\,U_{rt} \right]& r^2 \left[\cos \theta_k\,U_{rr} - U_{rt} \right] &  r^2 \left[\cos \theta_k\,U_{ri} - U_{ti} \right] \nonumber\\
r^2 \left[-\cos \theta_k\,U_{tt}+ U_{rt} \right]& r^2 \left[U_{rr} - \cos \theta_k\,U_{rt} \right] &r^2 \left[U_{ri} -\cos \theta_k\, U_{ti} \right] \\
U^j_{\;t} & U^j_{\;r} &U^j_{\;i}
\end{matrix}
\right)\,.
\eea
Here $U_{\mu\nu}=\partial_\mu(\delta u) \partial_\nu(\delta u)+\partial_\mu(\delta \theta) \partial_\nu(\delta \theta)$, and upper $j$ indices are raised with $g^{(S^4)}$.

Plugging (\ref{eq:delM}) into (\ref{eq:expansion}) obtains, up to quadratic order,
\bea\label{eq:expansion2}
\sqrt{-\det\,M} &=&  - \frac{\sin^3 \theta_k}{r^2} \delta u^2  - \frac{1}{2} \frac{\sin^3 \theta_k}{r^2} \left[r^2 \left(-(\partial_t \phi_a)^2+ (\partial_r \phi_a)^2\right) + \partial^i \phi_a \partial_i \phi_a  \right]+ \nonumber\\
& -& \frac{1}{r^2} \sin^3 \theta_k \cos \theta_k (r^2 f_{rt}+ 4 \sin \theta_k\,\delta \theta) + \frac{\sin \theta_k}{2r^2} \left[r^4 f_{rt}^2 -  f_{ij}^2 + r^2(-f_{ti}f^i_{\;t}+f_{ri}f^i_{\;r}) \right]+ \nonumber\\
&-& \frac{2}{r^2} (1+2 \cos 2 \theta_k) \sin^3 \theta_k \,\delta \theta^2-4 \sin^2 \theta_k \,\cos^2 \theta_k\,\delta \theta f_{rt}\,.
\eea
Here the kinetic term for $\phi_a$ stands for the sum over the kinetic terms for $\delta u$ and $\delta \theta$.

Next we consider the CS term; up to quadratic order, it gives
\bea\label{eq:CSexpansion}
F \wedge C_4 &=& d^6 \xi \sqrt{g^{(S^4)}} \left[4 \frac{\cos \theta_k}{r^2} \sin^4 \theta_k \delta \theta- \left(\frac{3}{2} \theta_k - \sin 2 \theta_k+ \frac{1}{8} \sin 4 \theta_k \right)f_{rt} \right. \nonumber\\
&-& \left. 4 \sin^4 \theta_k \,\delta \theta f_{rt} + \frac{8}{r^2} \cos^2 \theta_k\,\sin^3 \theta_k\,\delta \theta^2 \right]\,.
\eea

Finally, combining (\ref{eq:expansion2}) and (\ref{eq:CSexpansion}) obtains
\bea\label{eq:AfinalS}
S^{(2)} &=& T_5 \int d^6 \xi \sqrt{g^{(S^4)}} \left\lbrace \frac{\sin^3 \theta_k}{r^2} \frac{1}{2}  \left[r^2 (\partial_t \phi_a)^2 - (\partial_r \phi_a)^2 - \partial^i \phi_a \partial_i \phi_a - 2 \delta u^2 + 4 \delta \theta^2 \right] \right. \nonumber\\
&+& \left.\frac{\sin \theta_k}{2} \left(r^2 f_{rt}^2 - \frac{1}{2r^2} f_{ij}^2 - f_{ti} f^i_{\;t}+ f_{ri}f^i_{\;r} \right) - 4 \sin^2 \theta_k\,\delta \theta\,f_{rt} \right\rbrace\,.
\eea
This is our final 6d expression. (The term linear in $\delta \theta$ cancels, while the linear term in $f_{rt}$ is a total derivative.)

\subsection{Dimensional reduction}\label{app:dimred}

\subsubsection*{$SO(3)$ triplet}

We first consider the $SO(3)$ triplet; we just found that it decouples from the rest of the modes, and this is consistent with the $SO(3)$ symmetry. It is enough to consider the magnitude of the 3-vector, $\delta u$; it is expanded in spherical harmonics
\be
\delta u = \sum_l \delta u_l(\sigma) Y^l(y)\;\;,\;\;\nabla^2_{S^4}Y^l=-l(l+3) Y^l\,.
\ee
Here $\sigma$ are coordinates along $AdS_2$, $y$ are coordinates on $S^4$, and $l$ is a multi-index (see below). The action follows from (\ref{eq:AfinalS}):
\be
S \supset T_5 \sin^3 \theta\,\sum_{lm}\,\frac{1}{2} \int d^2 \sigma \frac{1}{r^2} \left(   r^2 (\partial_t \delta u_{lm})^2- r^2 (\partial_r \delta u_{lm})^2- \left(l(l+3)+2 \right) \delta u_{lm}^2 \right)\,.
\ee
Then the mass of the $l$-th KK mode of the $SO(3)$ triplet and the dual dimension are
\be
m_l^2 = 2 + l(l+3)\;,\;h_l=l+2\,.
\ee

\subsubsection*{Internal gauge field}

The 6d gauge field gives $a_\mu = (a_t, a_r, a_i)$, where $a_i$ are along $S^4$. We now consider these internal components, that reduce to scalars along $AdS_2$. In order to decouple them from $(a_t, a_r)$, we choose transverse gauge
\be
\nabla^i a_i =0\,.
\ee
Then, for instance,
$$
f_{ti} f^i_{\;t} \to  - \partial_t a_i \partial_t a^i - \partial_i a_t \partial^i a_t\,.
$$
The cross-term $\partial_i a_t \partial_t a^i$ has disappeared after integration by parts and use of the gauge fixing condition.

Then the action for $a_i$ simply becomes
\be
S \supset T_5 \sin \theta_k \,\int d^6 \xi\, \frac{1}{2}\left((\partial_t a_i)^2 - (\partial_r a_i)^2 + \frac{1}{r^2} a_i \left(g^{ij}_{(S^4)}\nabla^2_{(S^4)}- R^{ij}_{(S^4)}\right)a_j\right)
\ee
where $i,j$ indices are raised and lowered with $g_{(S^4)}$, and the curvature of the internal sphere comes from integrating by parts the term $f_{ij} f^{ij}$. For a unit-radius $S^n$, $R^{ij}=(n-1) g^{ij}$.

The KK reduction is performed by expanding $a_i$ in vector spherical harmonics on $S^4$. Let us review the construction of vector harmonics for $S^n$.\footnote{We thank N. Bobev for explanations on vector harmonics.} Start from the fundamental representation $Y^A$ of $SO(n+1)$, $A= 1,\ldots, n+1$, which satisfies
$$
\nabla^2 Y^A = - n Y^A\,.
$$
Spherical harmonics are gien by symmetric traceless products:
$$
Y^{A_1 \ldots A_l} \equiv Y^{(A_1} \ldots Y^{A_l)}\;\;,\;\;\nabla^2 Y^{A_1 \ldots A_l}= - l (l+n-1) Y^{A_1 \ldots A_l}\,.
$$
These are the harmonics that were denoted by $Y^l$ above.

Next we introduce
\be
Y_i^A \equiv \partial_i Y^A\,.
\ee
Some useful properties are
\be
\sum_i Y_i^A Y_i^B = -Y^A Y^B + \delta^{AB}\;\;,\;\;\sum_A (Y^A)^2 =1\;\;,\;\;\nabla_i Y_j^A = - \delta_{ij}Y^A\,.
\ee
Notice that $Y_i^A$ is an eigenvector of the laplacian, but it is not transverse. 

The transverse vector harmonics are constructed as
\be
Y_i^{A_1 \ldots A_l} = Y_i^{[A_1} Y^{(A_2]}\ldots Y^{A_l)}\;\;,\;\;l \ge 2\,,
\ee
where the first two indices are antisymmetrized and the remaining $l-1$ ones are symmetrized.
This gives a representation of $SO(n+1)$ with $l-1$ boxes in the first row and one in the second --this will be denoted by $(2,l-2)$. The calculation of the eigenvalues gives
\be\label{eq:vec-eigenv}
\left(\nabla^2_{S^n}-(n-1)\right) Y_i^{A_1 \ldots A_l} = - l(l+n-3) Y_i^{A_1 \ldots A_l}
\ee

Now we are ready to perform the dimensional reduction, specializing the above results to $n=4$. Expanding
\be
a_i(\sigma, y ) = \sum_l a_l(\sigma) Y_i^l(y)
\ee
(where $l$ is a multi-index standing for $A_1, \ldots ,A_l$) we find
\be
S \supset\frac{1}{2}T_5 \sin \theta_k \,\sum_l\int d^6 \xi\, \frac{1}{r^2}\left(r^2(\partial_t a_l)^2 - r^2(\partial_r a_l)^2 - l (l+1) a_l^2\right)\;\;,\;\;l \ge 2\,.
\ee
So the $(2,l-2)$ vector harmonic has a mass $m_l^2 = l (l+1)$
and is dual to an operator of dimension $h_l=l+1$.

\subsubsection*{$(\delta \theta, f_{rt})$ sector}

In this sector the action is
\bea
S&\supset& T_5 \int d^6 \xi \sqrt{g_4} \left \lbrace \sin^3 \theta_k \left(\frac{1}{2} (\partial_t \delta \theta)^2 - \frac{1}{2} (\partial_r \delta \theta)^2 -\frac{1}{2r^2} (\partial_i \delta \theta)^2 + \frac{2}{r^2} \delta \theta^2\right) \right. \nonumber \\
&+& \frac{\sin \theta_k}{2}r^2 \left((\partial_r a_t)^2 + (\partial_t a_r)^2 -2 \partial_r a_t \partial_t a_r + \frac{1}{r^2} (\partial_i a_t \partial^i a_t -  \partial_i a_r \partial^i a_r )\right) \nonumber\\
&-& \left. 4 \sin^2 \theta_k\,\delta \theta (\partial_r a_t - \partial_t a_r)\right \rbrace\,.
\eea

We still have gauge redundancies along the 2d space, and we use them to set $a_0=0$, \textit{after} imposing Gauss' law:
\be
\partial_r (r^2 \partial_t a_r + 4 \sin \theta_k\,\delta \theta) =0\,.
\ee
Then the equations of motion simplify to
\bea
&& r^2 (- \partial_t^2 \delta \theta+\partial_r^2 \delta \theta) + \nabla_{S^4}^2 \delta \theta+ 4 \delta \theta + 4 \frac{r^2}{\sin \theta_k} \partial_t a_r=0\nonumber\\
&&- \partial_t (r^2 \partial_t a_r + 4 \sin \theta_k\,\delta \theta)+ \nabla_{S^4}^2 a_r=0\,.
\eea
Defining the combination
\be
\eta \equiv \frac{1}{\sin \theta_k} r^2 \partial_t a_r + 4 \delta \theta
\ee
and expanding in spherical harmonics, obtains
\bea\label{eq:systemtheta}
&& \partial_r \eta_l=0\nonumber\\
&& r^2 (- \partial_t^2 \delta \theta_l+\partial_r^2 \delta \theta_l) - l(l+3) \delta \theta_l- 12 \delta \theta_l + 4 \eta_l=0\nonumber\\
&&- r^2 \partial_t^2 \eta_l - l(l+3) (\eta_l -4 \delta \theta_l)=0\,.
\eea
The last equation here follows by taking one time derivative on the $a_r$ equation above.

For the $l=0$ mode, this system gives $\partial_t \eta_0 = \partial_r \eta_0 = 0$, so that $\eta_0 = const$. Then the equation for $\delta \theta$ gives  $m_0^2 = 12$, which is dual to an operator of dimension $h_0=4$.
For the higher $l \ge 1$ modes, diagonalizing the mass matrix in (\ref{eq:systemtheta}) obtains
\be
m_l^2 = \left \lbrace 
\begin{matrix}
(l+3)(l+4) \\
l(l-1)
\end{matrix}
 \right.\;\;\Rightarrow\;\;h_l =
 \left \lbrace 
\begin{matrix}
l+4 \\
l
\end{matrix}
 \right.
\ee
in agreement with \cite{Paredes}. This ends our analysis of the D5 fluctuations.

\end{document}